\documentclass[reprint,amsmath,amssymb,aps]{revtex4-1}
\usepackage{graphicx}
\usepackage{dcolumn}
\usepackage{bm}
\usepackage{float}
\usepackage{epsfig,color,amsmath}
\usepackage{mathrsfs}

\begin{document}

\title{Target finding in fibrous biological environments}
\author{David Gomez$^{1}$, Eial Teomy$^{1}$, Ayelet Lesman$^{1}$, Yair Shokef$^{1,2,3}$}
\affiliation{$^{1}$School of Mechanical Engineering, Tel Aviv University, Tel Aviv 69978, Israel}
\affiliation{$^{2}$Sackler Center for Computational Molecular and Materials Science, Tel Aviv University, Tel Aviv 69978, Israel}
\affiliation{$^{3}$Kavli Institute for Theoretical Physics, University of California, Santa Barbara, California 93106, USA}

\begin{abstract}
We {\color{black}use a lattice model to} study first-passage time distributions of target finding events through complex environments with elongated fibers distributed with different anisotropies and volume occupation fractions. For isotropic systems and for low densities of aligned fibers, the three-dimensional search is a Poisson process with the first-passage time exponentially distributed with the most probable finding time at zero. At high enough densities of aligned fibers, elongated channels emerge, reducing the dynamics dimensionality to one dimension. We show how the shape and size of the channels modify the behavior of the first-passage time distribution and its short, intermediate, and long time scales. We develop an exactly solvable model for synthetic rectangular channels, which captures the effects of the tortuous local structure of the elongated channels that naturally emerge in our system. {\color{black}For arbitrary values of the nematic order parameter of fiber orientations, we develop a mapping to the simpler situation of fully aligned fibers at some other effective volume occupation fraction.} Our results shed light on the molecular transport of biomolecules between biological cells in complex fibrous environments.
\end{abstract}

\maketitle

\section{\label{sec:Introduction}Introduction }

Many biochemical reactions between chemically active molecules involve molecules distant in space, and commonly, at least one molecular species is free and searches for its target molecule. Thus, biochemical reactions depend on how a molecule diffuses toward its target, and \textcolor{black}{also on} the probability associated with the molecules to react once they are in close contact~\cite{Calef, Weiss, Smoluchowski}. The former process depends on the reactant's diffusion coefficient $D$ and the shape and size of the confining region. The latter process depends on an intrinsic reactivity $k$. Under ideal conditions, reactants in high concentrations are uniformly distributed in space, leading to uniform and independent encounters between molecules. Consequently, diffusion and kinetic controls are each correctly described by a single time-scale, and in particular, the mean reaction time is given as the sum of two time-scales: the mean time for molecular encounter $\sim 1/D$, and the mean time for chemical reaction $\sim 1/k$~\cite{Collins, Berg, North}. In recent years it has become more evident that it is necessary to question this simplified description of biochemical reactions and consider more elaborate models. In particular, for many biochemical reactions, the number of reactants can be low, limited to a few copies~\cite{Guptasarma}. For example, gene expression and gene regulation occur at low-copy protein numbers, and their stochastic behavior has been the focus of many studies~\cite{Paulsson01, Maheshri, Morelli, Brenner}. Another example is the sensory systems used by swimming bacteria responding to the activation-deactivation of membrane receptors by a limited amount of molecules~\cite{Berg2, Sourjik, Kaizu}. In these cases, it is no longer appropriate to describe reaction rates with the mean time for a molecular encounter or the mean first-passage time (MFPT), but one needs to know the whole distribution of first-passage times (FPTs). It becomes then clear that at low molecular concentrations, the MFPT is not the only relevant time scale of the reaction process, but the most-probable FPT (MPFPT) becomes essential too. Hence, different diffusive-controlled events in the same system can vary widely in their time scales since the MPFPT, and the MFPT can differ by orders of magnitude~\cite{Godec, Grebenkov}. \textcolor{black}{Knowledge of the whole distribution is needed also to determine the record statistics~\cite{Majumdar2010, Wergen2013, Hartich2019}, the statistics of multi-particle systems~\cite{Bray2013, Lawley2020, Lawley2020b, Madrid2020, Hartich2018, Hartich2019b, Mejia2011}, and to identify intermediate states in transition networks~\cite{Li2013, Thorneywork2020}.}

Moreover, biological systems typically encompass complex environments in which molecules with different shapes, sizes, and chemical compositions coexist~\cite{Minton, Han2, Kim2}. The crowded nature of biological systems has motivated great efforts to understand its effects on enzymatic activity, protein folding, and gene regulation~\cite{Zhou, Ellis, Gomez1, Gomez2, Gomez3}. Additionally, in highly dynamical environments such as the extracellular matrix (ECM) of living tissues, the geometric structure is continuously remodeled by cellular activities such as traction forces, degradation, or secretion of ECM fibers~\cite{Kim, Trubelja, Notbohm1, Kielty, Page-McCaw, Wang1, Han, Spill, Iozza, Schwager, Jansen}. Such ECM remodeling modulates the fiber volume occupation fraction and the anisotropy of the fibers. These, in turn, can dramatically affect the diffusion of molecules through the matrix~\cite{Trubelja, Frantz, Gomez4, Jung, Mann, Goren, Ban, Stopak, Vader, Kim4}.

We have recently shown that increased density and alignment of fibers facilitate molecular transport from a source to a target, which may support long-range cell-cell biochemical interactions~\cite{Gomez4}. In our {\color{black}3D lattice} model (see Fig.~\ref{model}A), we consider random walks of a diffusing molecule searching for its target within a system with fixed elongated fibers distributed with a nematic order parameter $S$ and taking up a volume fraction $\phi$. Only excluded volume interactions are considered. As the density of fibers increases, the system percolates differently depending on the alignment: for isotropic fibers ($S=0$), the system undergoes a drilling percolation transition at $\phi_C^{3D}=0.75$~\cite{Kantor, Schrenk, Grassberger}. At fiber occupation fractions of $\phi \geq \phi_C^{3D}$, the molecule gets caged by fibers, impeding target finding. Whereas for aligned fibers ($S=1$), the system follows a 2D random site percolation process in the cross section of fiber positioning, with a critical fiber density of $\phi_C^{2D}=0.408$~\cite{Stauffer}. As $\phi$ reaches $\phi_C^{2D}$, the components of the diffusion coefficient perpendicular to fiber alignment decay to zero while the parallel component remains unaffected. This effect on the diffusion coefficient results from the emergence of channel-like structures that confine the dynamics to a 1D process, which is more effective than the 3D case. This caged state of the dynamics can be modulated toward the 1D process by continuously increasing fiber alignment $S$ and fiber volume fraction $\phi$.

In this paper, we study the effect of channel shape and size on the FPT \textcolor{black}{probability density}, the MFPT, and the MPFPT of a target-finding process by numerical simulations and by analytically solving for several simplified geometries of channels with square and rectangular cross sections. We show that the channel size as well as its fractal shape influence the FPT \textcolor{black}{probability density}. Additionally, we show that the FPT \textcolor{black}{probability density} is no longer characterized by a single time-scale, implying that the typical notion of describing molecular reactions as the sum of two MFPTs ($1/D$ and $1/k$) is not appropriate. \textcolor{black}{We also consider intermediate fiber alignment values, between the isotropic to the fully aligned. Interestingly, we construct a mapping between this more complex case to the simpler case described above of fully aligned fibers, and find that this mapping is effective in describing both the MFPT and the FPT probability density.}

The paper is organized as follows: in Sec.~\ref{Computational}, we introduce our model. In Sec.~\ref{3DMain}, we study the FPT problem in the free case ($\phi=0$) and in the low fiber volume occupation fraction regime, both for aligned ($S=1$) and for isotropic ($S=0$) fiber distributions. Next, in Sec.~\ref{1DMain} we consider the very high volume occupation fraction limit ($\phi \approx 1$) of aligned fibers ($S=1$), such that the dynamics are entirely 1D. Then, in Sec.~\ref{Quasi} we consider high densities ($\phi>\phi_C^{2D}$) of aligned ($S=1$) fibers and study the FPT in elongated channels with different shapes and sizes: we numerically analyze the FPT \textcolor{black}{probability densities} that occur in the disordered channels that naturally emerge in our system of aligned fibers, and we approximate these complex channels with synthetic square and rectangular channels that allow us to study the FPT \textcolor{black}{probability density} analytically. In Sec.~\ref{Intermediate}, we study the FPT problem for intermediate alignment $0<S<1$, and map these more general cases to our results for perfectly aligned fibers ($S=1$), both above and below the percolation threshold. Finally, Sec.~\ref{Discussion} concludes with a summary and discussion of our work and its implications. 

\section{\label{Computational}Computational model}

\begin{figure}
\centering
\includegraphics[width=\columnwidth]{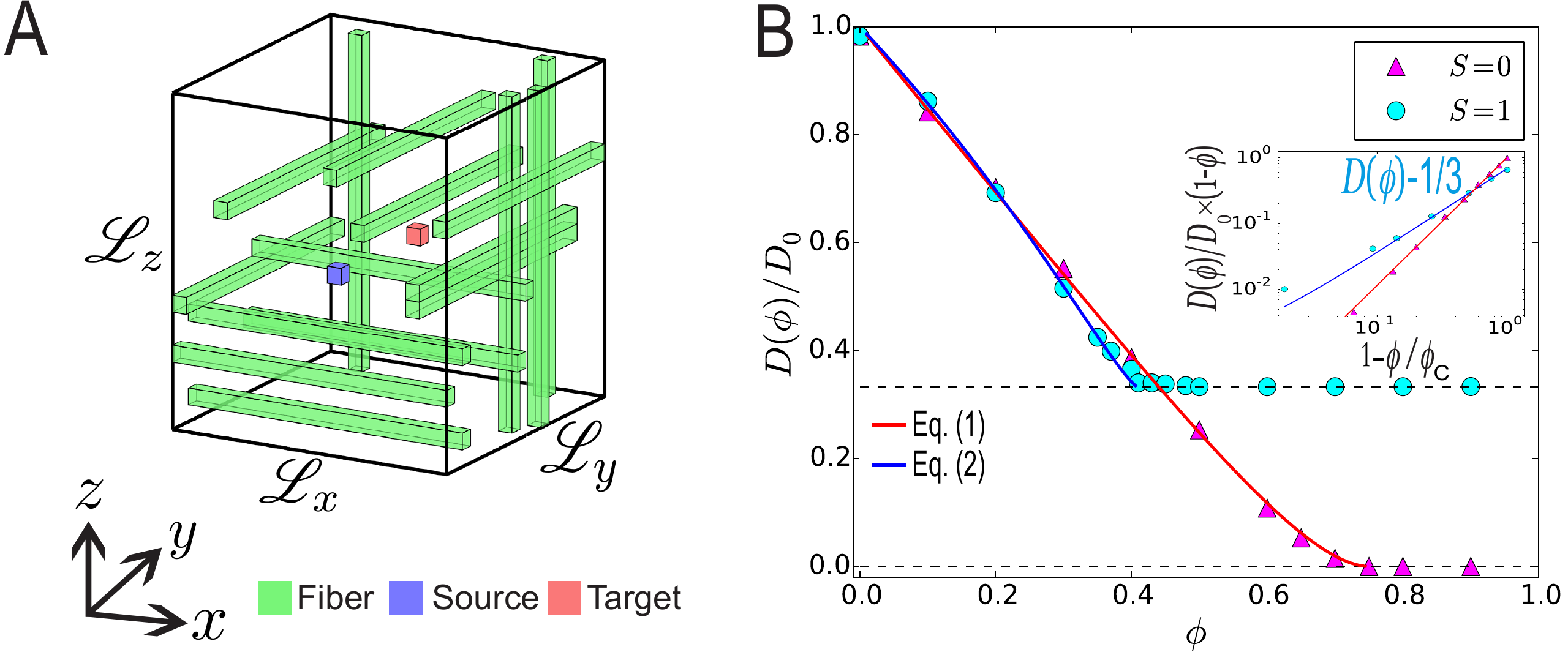} 
\caption{{\color{black}A) Schematic representation of our lattice model. B) Fiber density modulates the diffusion coefficient. The ratio $D(\phi)/D_0$ decreases as a function $\phi$. Inset: log-log plot of $D(\phi)/D_0 \times (1-\phi)$ as a function of $(1-\phi/\phi_C)$. For $S=1$, we consider only diffusion along the $x-z$ plane. Therefore we remove a value of $1/3$, which accounts to the contribution in the diffusion coefficient along the $y$ axis.}}
\label{model}
\end{figure}

We study transport in complex environments using a model of particles moving on a 3D simple cubic lattice with periodic boundary conditions in all directions. The lattice is set with a total volume of $V= {\mathscr L}_x  \times {\mathscr L}_y \times {\mathscr L}_z$ sites, and three types of molecules can occupy the different lattice sites: a tracer molecule that is released from a source located at $\textbf{r}_0=(x_0,y_0,z_0)$, with $ x_0=\left\lfloor{\mathscr L}_x/2\right\rfloor$, $y_0=\left\lfloor{\mathscr L}_y/4\right\rfloor$, $z_0=\left\lfloor{\mathscr L}_z/2\right\rfloor $, a static target placed at $\textbf{r}_T=(x_T,y_T,z_T)$, with $x_T=\left\lfloor{\mathscr L}_x/2\right\rfloor$, $y_T=3\left\lfloor{\mathscr L}_y/4\right\rfloor$, $z_T=\left\lfloor{\mathscr L}_z/2\right\rfloor$, and elongated fibers, each one running along one of the three principal directions of the lattice and that span the whole system length, {\color{black} see Fig.~\ref{model}A.} \textcolor{black}{We consider fibers with thickness of one lattice site and allow them to cross each other. Each fiber has a probability $p_i$ of being oriented along the $i=x,y,z$ axis. We calibrate the probabilities $p_i$ to obtain a desired nematic order parameter of the system $S=(3 \langle \cos^{2} \theta \rangle -1)/2 $, with $\theta$ the angle between the fiber orientation and the preferred direction of orientation~\cite{Mercurieva}, which we choose to be the $y$-axis. Note that the preferred direction of fiber orientation and the line connecting the source and target are the same. The motivation for this choice is that cells stretch the matrix and cause fibers to orient along the direction between the cells~\cite{Gomez4}. Therefore, we set $p_x$ to be equal to $p_z$, thus $S=(3p_y - 1)/2$, and $p_x = p_z = (1-p_y)/2$. A system with $S=0$ has fibers isotropically distributed, with $p_x=p_y=p_z=1/3$, and one with $S=1$ has all the fibers aligned along the $y$-axis, i.e., $p_x=p_z=0$ and $p_y=1$. Thus, for a system with given values of $S$ and of the total volume fraction $\phi =1 - \left[1 - Mp_{x}/({\mathscr L}_y {\mathscr L}_z) \right] \left[1 - M p_{y}/({\mathscr L}_x {\mathscr L}_z) \right] \left[1 - M p_{z}/({\mathscr L}_x {\mathscr L}_y) \right]$, we compute the total number of fibers $M$, and distribute them in the lattice with probabilities $p_i$.} The FPT is defined as the time needed for the tracer molecule to reach the target site for the first time. After every finding event, the tracer is placed back to the source location, and then, a new random configuration of fibers is generated, and a new target search process begins.  

\textcolor{black}{When an exponentially distributed clock with mean $1$ ticks, the molecule advances to one of its six neighboring sites, provided the desired location is empty of fibers. Otherwise, the move is rejected. Whether the molecule moved or not, the clock resets. In the simulations, the molecule attempts to move at constant time steps. The difference between the described model and the simulation results appear only at very short times, which are not analyzed here.} 

For the case $\phi=0$, with no fibers, the molecule moves in an empty lattice and diffuses with a diffusion coefficient $D_0= 1/6$. As the volume occupation fraction increases toward the percolation threshold $\phi_C$, the diffusion coefficient decays algebraically as predicted by percolation theory~\cite{Stauffer}. Hence, to describe the complex behavior of the diffusion coefficient one needs the critical density $\phi_C$ and the exponent $\mu$, which controls this algebraic decay~\cite{Novak}. In the isotropic ($S=0$) case, diffusion is equally hindered in all directions and the diffusion coefficient decays following the Swiss-cheese model~\cite{Novak}, as: 
\begin{equation}
\frac{D(\phi)}{D_0} = \frac{\left( 1- \frac{\phi}{ \phi_{C}^{3D} } \right)^{\mu_{0} } } {\left(1-\phi\right)}. \label{DifIoso}
\end{equation}
Using the critical density for drilling percolation $\phi_C^{3D}=0.75$, we obtain $\mu_{0}=2$, agreeing with the value reported in~\cite{Stauffer}, see Fig.~\ref{model}B. For aligned fibers ($S=1$), diffusion is not affected along the $y$ axis, but is hindered along the $x-z$ cross section. Thus, we decouple the effect of fibers in the components of the diffusion coefficient, and see that it decays as: 
\begin{equation}
\frac{D(\phi)}{D_0}  = \frac{1}{3} + \frac{2}{3} \times \frac{\left(1- \frac{\phi}{ \phi_{C}^{2D}}\right)^{\mu_{1}} } { \left(1-\phi\right)}, \label{DifAlig}
\end{equation}
with the critical density for 2D random site percolation $ \phi_C^{2D}=0.408$~\cite{Stauffer}. We get that for our system $\mu_{1}=1.3$, in agreement with the value reported in~\cite{Stauffer}, see Fig.~\ref{model}B\textcolor{black}{, where we plot $D(\phi)$ and its algebraic scaling for $S=0$ and for $S=1$}.

\begin{figure}
\centering
\includegraphics[width=0.65\columnwidth]{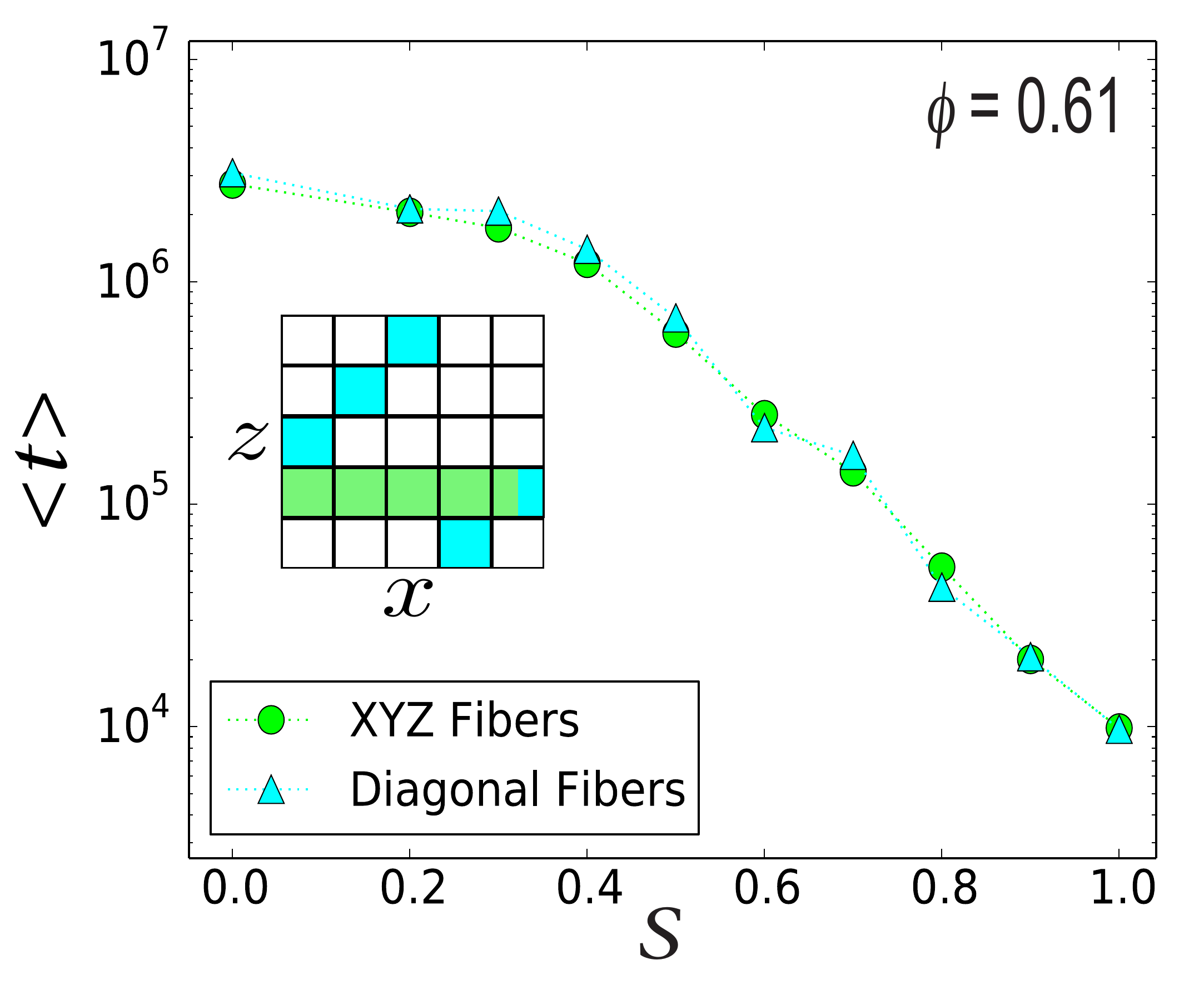} 
\caption{\textcolor{black}{MFPT as a function of nematic order parameter $S$ with diagonal (blue triangles) and straight (green circles) fibers at a fiber volume occupation fraction of $\phi=0.61$. Inset: schematic of diagonal (blue) and straight (green) fibers running across the $x-z$ plane.}}
\label{Diagonal}
\end{figure}

\textcolor{black}{Our discrete lattice model imposes limitations on fiber geometry, and the results obtained for it clearly differ from those for models with continuous positions and orientations of fibers. To get a sense of the effect of the discrete possible orientations that fibers can take in our model, we extend our model to allow fibers to also run in the main diagonal directions of the lattice. We compute the MFPT with and without diagonal fibers at different values of the nematic order parameter $S$ and at a fixed value of the fiber volume occupation fraction. Figure~\ref{Diagonal} presents the very good agreement between the two setups. Thus, for the remainder of our work, we exclude diagonal fibers and obtain intermediate values $0<S<1$ of the nematic order parameter of fiber orientations only by changing the fractions $p_x$, $p_y$ and $p_z$ of fibers along the three principal directions of the lattice.}

\section{\label{3DMain}3D Searching}

\begin{figure*}
\centering
\includegraphics[width=\textwidth]{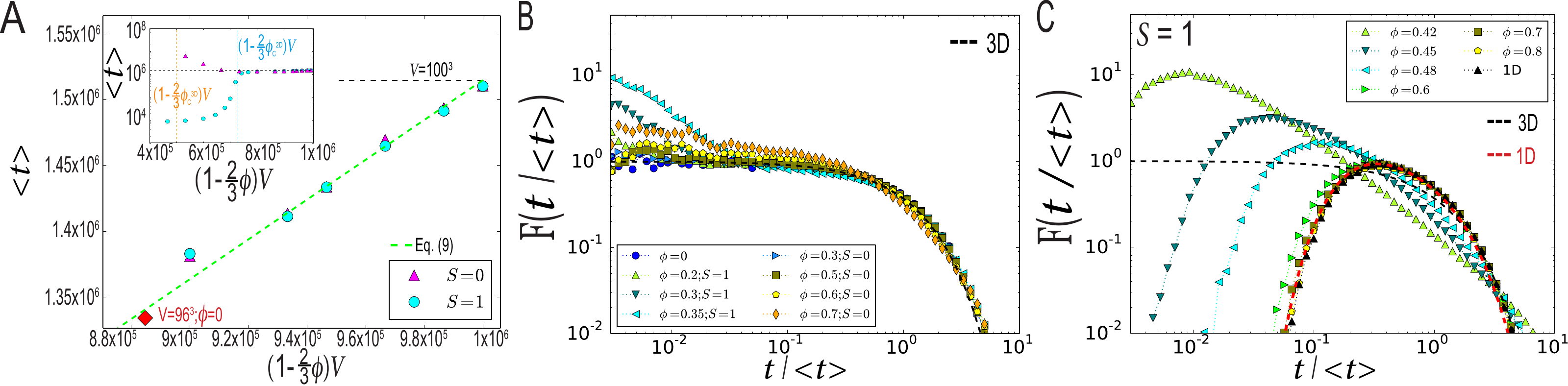} 
\caption{\label{fig:wide}MFPT and FPT \textcolor{black}{probability densities} for target finding for different values of $S$ and $\phi$. A) MFPT as a function of the available volume for $S=0$ and $S=1$. In the free case $\phi=0$, the MFPT increases with the volume. For low $\phi$, the MFPT is given by Eq.~(\ref{mfpt_lowPhi}) (lime line). Inset: The target finding dynamics strongly changes as $\phi$ approaches $\phi_C^{2D}$ (blue line) and $\phi_C^{3D}$ (orange line), for $S=1$ and $S=0$, respectively. B) Normalized FPT \textcolor{black}{probability density} for $S=0$ and $S=1$ for values of $\phi \leq \phi_C^{3D}$ and $\phi \leq \phi_C^{2D}$, respectively. Black dashed line shows exponential behavior. C) Normalized FPT \textcolor{black}{probability densities} for aligned fibers and values of $\phi >\phi_C^{2D}$. The red line is Eq.~(\ref{1DFPT}) for the 1D FPT \textcolor{black}{probability density}. Statistics are performed for $2\times10^{5}$ finding events.}
\label{fig1}
\end{figure*}

We start by considering the reference case of target finding without fibers ($\phi = 0$). This system with periodic boundary conditions allows us to solve the FPT \textcolor{black}{probability density}, based on the evolution equation for the probability $P(\textbf{r},t)$ of the particle to be at position $\textbf{r}=(x,y,z)$ at time $t$ \textcolor{black}{given that it has not reached the target yet}. The FPT \textcolor{black}{probability density} $F(\textbf{r}_T,t)$ of finding the target located at $\textbf{r}_T$ at time $t$, given that the molecule started at $\textbf{r}_0$ at $t=0$, is related to $P(\textbf{r},t)$ by~\cite{Siegert1951}:
\begin{align}
P\left(\textbf{r}_T,t\right)= \int^{t}_{0}P\left(\textbf{r}_0,t-t'\right)F\left(\textbf{r}_T,t'\right)dt' .\label{fs_derText}
\end{align}
The right hand side of Eq.~(\ref{fs_derText}) is the probability of reaching $\textbf{r}_T$ for the first time, given that at any previous time $t-t'$ the molecule was at $\textbf{r}_0$ \textcolor{black}{and has not yet visited the site $\textbf{r}_{T}$}. After taking the Laplace transform of both sides, we get the relation:
\begin{align}
\tilde{F} \left(\textbf{r}_T,s\right)=\frac{\tilde{P}\left(\textbf{r}_T,s\right)}{\tilde{P}\left(\textbf{r}_0,s\right)}. \label{eq1Text}
\end{align}
By separation of variables, $P\left(\textbf{r},t \right)=P_{\textcolor{black}{X}}\left(x,t\right)P_{\textcolor{black}{Y}}\left(y,t\right)P_{\textcolor{black}{Z}}\left(z,t\right)$, we obtain the independent probabilities $P_{\textcolor{black}{J}}(j,t)$, and find that the MFPT scales with the system's volume~\cite{Condamin2005, Condamin2007, Benichou2008, Guerin2016, Benichou2018}, see Appendix~\ref{3D}:
\begin{align}
\left\langle t\right\rangle = \frac{ \alpha_0 V}{2D_0} ,\label{mfpt_asymTex}
\end{align}
where~\cite{Condamin2005}
\begin{align}
\alpha_0=\int^{1}_{0}\int^{1}_{0}\int^{1}_{0}\frac{dxdydz}{\Omega_{0}(x,y,z)}\approx0.505 \label{mfpt_appText}
\end{align} 
is a geometrical prefactor, and $\Omega_{0}\left(x,y,z\right)=\omega\left(\pi x\right)+\omega\left(\pi z\right)+\omega\left(2\pi y\right)$ with the function $\omega(\psi)=(1-\cos \psi) / 3$. Here, the periodic boundary conditions ensure that over long times, the tracer molecule is equally likely to be at any lattice site in the system, making the finding events a Poisson process, with an exponential FPT \textcolor{black}{probability density} $F(t)=1/\left\langle t\right\rangle \times \exp \left( -t/\left\langle t\right\rangle \right) $~\cite{Kingman}. To test the theoretical prediction of Eq.~(\ref{mfpt_asymTex}), we plot in Fig.~\ref{fig1}A the MFPT for two system sizes with ${\mathscr L}_x={\mathscr L}_y={\mathscr L}_z=100$ (black dashed line) and ${\mathscr L}_x={\mathscr L}_y={\mathscr L}_z=96$ (red diamond). We recover the MFPT dependence on the system's volume and the free diffusion coefficient $D_0$, as shown by the green dashed line with $\phi=0$.  

Next, we examine the case of low values of fiber occupation fraction $\phi$ and their effect on the MFPT. The introduction of fibers has two competing effects on the MFPT. On the one hand, as $\phi$ increases, the available volume decreases, reducing the MFPT for target finding. On the other hand, the presence of fibers hinders diffusion, thus increasing the time needed for the tracer to find its target. \textcolor{black}{It is conjectured~\cite{Grebenkov2017} that the second effect prevails, i.e. that adding fibers increases the MFPT.} In our model, we capture the MFPT behavior by using a mean-field approximation for the case $\phi \ll 1$. Specifically, for aligned fibers ($S=1$), diffusion is hindered only along the $x-z$ plane, and we approximate by $1-\phi$ the probability of succeeding to move in this plane. For other values of $S$, the diffusion is also hindered in the $y$ axis. Hence, we approximate by $1-\phi\left(p_{x}+p_{y}\right)=1-\phi\left(p_{z}+p_{y}\right)=1-  \phi(2+S)/3$ the probability of succeeding to move in the $x-z$ plane, and by $1-\phi\left(p_{x}+p_{z}\right)=1-2\phi(1-S)/3$ the probability of succeeding to move in the $y$ axis. Therefore, the function $\Omega_{0}\left(x,y,z\right)$ used in Eq.~(\ref{mfpt_appText}) for $\phi=0$ depends on $\phi$ and on $S$ via the new function 
\begin{align}
&\Omega\left(x,y,z,\phi,S\right)=\left(1-\phi\frac{2+S}{3}\right)\left[\omega\left(\pi x\right)+\omega\left(\pi z\right)\right]+\nonumber\\
&\left(1-2\phi\frac{1-S}{3}\right)\omega\left(2\pi y\right).
\end{align}
We suggest that $\alpha_0$ in Eq.~(\ref{mfpt_asymTex}) should be replaced by $\alpha(\phi)$ which is obtained by substituting this expression for $\Omega_{0}$ in Eq.~(\ref{mfpt_appText}). After expanding $\alpha(\phi)$ to first order in $\phi$, we obtain that $\alpha(\phi) =\alpha_0 -\alpha_1\phi$, with 
\begin{align}
\alpha_1= \int^{1}_{0}\int^{1}_{0}\int^{1}_{0}\frac{\omega\left(\pi x\right)+\omega\left(\pi z\right)}{\Omega^{2}_{0}(x,y,z)}dxdydz=\frac{2}{3}\alpha_{0}\approx 0.337. 
\end{align}
Note that by definition $p_{x}+p_{y}+p_{z}=1$, and thus to first order in $\phi$, the MFPT is independent of $S$. Thus we expect that for low $\phi$ the MFPT will be given by
\begin{align}
\langle t \rangle = \left( 1-\frac{2\phi}{3}  \right)\frac{\alpha_0 V}{2 D_0}. \label{mfpt_lowPhi}
\end{align}
To test our predictions, we plot in Fig.~\ref{fig1}A the MFPT as a function of $(1-2\phi/3) V$ for \textcolor{black}{small} values $\phi<0.15$ of fiber volume occupation. We see that for these low values of $\phi$, fiber alignment $S$ does not affect target finding (circle and purple data), and \textcolor{black}{our theoretical prediction in Eq.~(\ref{mfpt_lowPhi}) describes well the numerical results}.

To characterize the target finding dynamics, we obtain the FPT \textcolor{black}{probability densities} for values of fiber occupation fractions below the percolation thresholds. \textcolor{black}{In the absence of fibers ($\phi=0$), the FPT probability density is very close to exponential, while for increasing values of $\phi<\phi_C$, deviations from exponential behavior appear at short times, see Fig.~\ref{fig1}B. Thus,} as the system approaches percolation, the FPT \textcolor{black}{probability densities} are not fully characterized by a single time scale. \textcolor{black}{Note that also without fibers we expect deviations from exponential at short times, as will be discussed in Sec.~\ref{sec:synth} below. However, for the system size shown here, these deviations occur at very short times that are beyond the range plotted in Fig.~\ref{fig1}B.}

Target finding dynamics for fiber densities higher than the critical thresholds are very different if fibers are aligned or isotropically distributed. In the case $S=0$, the diffusion coefficient decays to zero for $\phi > \phi_C^{3D}$, and thus, the MFPT diverges, and the FPT \textcolor{black}{probability density} is no longer defined, as shown in the inset of Fig.~\ref{fig1}A. For $S=1$, the MFPT follows a complex behavior for $\phi \geq \phi_C^{2D}$, exhibiting a sharp decrease around $\phi_C^{2D}$, as shown by the inset in Fig.~\ref{fig1}A. Moreover, the FPT \textcolor{black}{probability density} for $\phi \geq \phi_C^{2D}$ follows a non-monotonic behavior characterized by three different time scales: the MPFPT at short FPTs, the MFPT at intermediate time scales, and the time scale of distribution tail (TSDT), at long FPT, as shown in Fig.~\ref{fig1}C. We define the TSDT as the decay rate $\hat{t}$ of the exponential tail of the FPT \textcolor{black}{probability density}. For these cases, channels in the $x-z$ cross section are formed, reducing the dimensionality of the dynamics from 3D to 1D. \textcolor{black}{In Sec.~\ref{Quasi}, we study in detail how the channel structure affects the FPT probability density for fully aligned fibers ($S=1$), and in Sec.~\ref{Intermediate} we extend that to intermediate alignment $0<S<1$, but before these steps, in Sec.~\ref{1DMain} we first} consider the simpler full 1D limit, which is obtained for $\phi \approx1$ \textcolor{black}{and $S=1$}.

\section{\label{1DMain}1D searching}

\begin{figure*}
\centering
\includegraphics[width=0.65\textwidth]{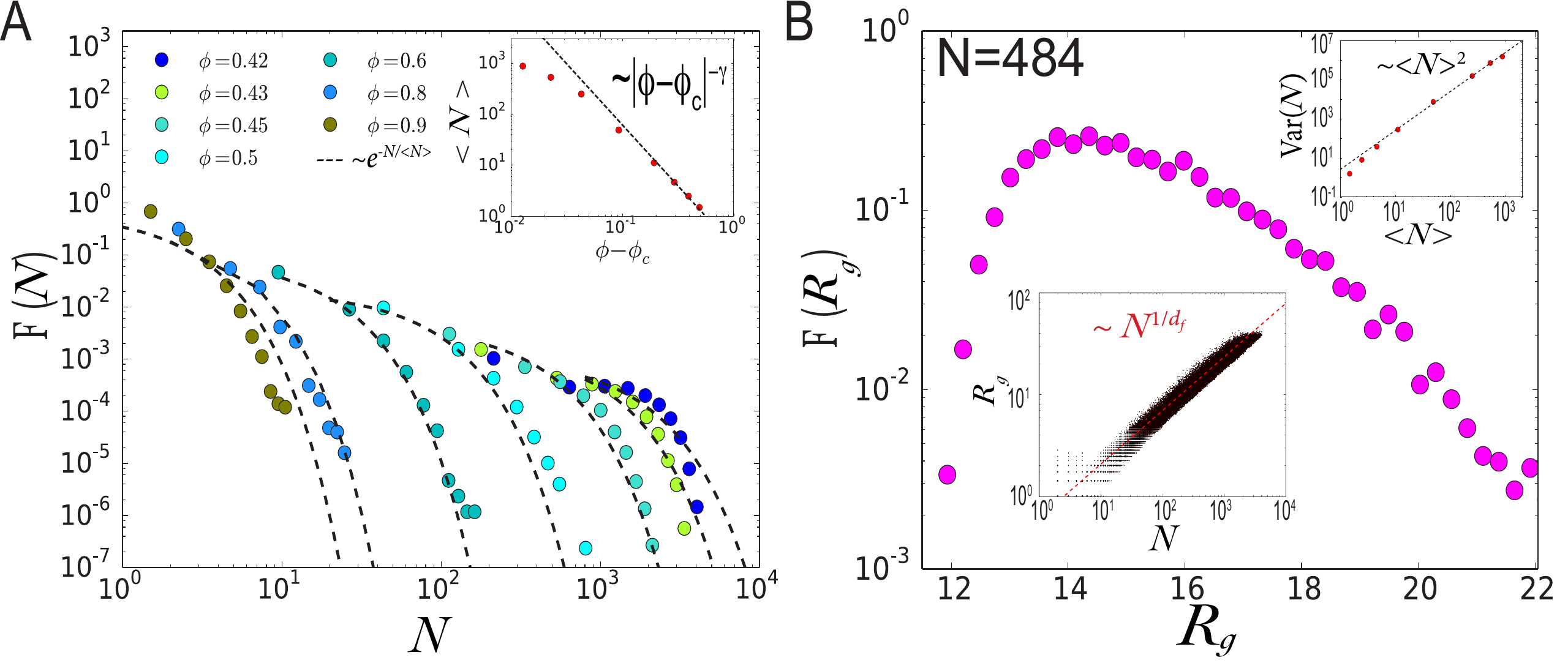} 
\caption{\label{fig:wide}Channel characterization for $S=1$. A) {\color{black}Probability density} of the $x-z$ cross-sectional area for different values of $\phi$. Black dashed lines are the exponential decay of the channel size distribution. Inset: the average channel size scales as $\mid \phi - \phi_C^{2D} \mid ^{-\gamma}$, with $\gamma=43/18$~\cite{Stauffer}. B) {\color{black}Probability density} of the radius of gyration of $4\times10^{5}$ channels with a fixed value of $N=484$. Lower inset:  The radius of gyration increases as $ N^{1/d_f}$. Upper inset: The variance of $N$ increases quadratically as a function of the mean channel size $\langle N\rangle$. Statistics are performed for $5\times10^{4}$ randomly generated fiber configurations for each volume fraction.}
\label{fig2}
\end{figure*}

In this section we present the liming case of a high density ($\phi \approx 1$) of aligned fibers ($S=1$), for which the diffusing molecule is confined to a 1D line along the direction of the fibers. In our lattice model, the channel is aligned along the $y$ axis and has a cross section equal to one. Due to the periodic boundaries of the system, the topology of the channel can be understood as a ring-like structure with a circumference of ${\mathscr L}_y$ and a single target that can be reached by the tracer molecule either from the left or from the right side of the ring. The {\color{black}probability density} of FPT to the target is related to the survival probability $\mathcal{H}(t)$ that the tracer did not yet reach the target up to time $t$ by~\cite{Klafter}: 
\begin{equation}
F(t) = - \frac{\partial  \mathcal{H}(y,t)} {\partial t}.
\end{equation}

Due to the periodic boundary conditions, the survival probability is equal to the probability that a tracer diffusing on a finite system of length ${\mathscr L}_y$ with absorbing boundary conditions \textcolor{black}{at $y=0$} remains in the system:
\begin{equation}
\mathcal{H}(t) =  \sum_{y=1}^{\mathscr{L}_{y}-1} P(y,t).
\end{equation}

We obtain the probability of finding the tracer at position $y$ at time $t$, $P(y,t)$, by solving the diffusion equation. In Appendix~\ref{1D} we show that for the specific case that the initial position of the tracer is equidistant from the two boundaries, i.e. $y_0={\mathscr L}_y/2$, the FPT \textcolor{black}{probability density} is given by:
\begin{align}
& F(t) = \frac{2}{{\mathscr L}_y}\sum^{{\mathscr L}_y/2}_{m=1}\omega\left(k_{2m-1}\right) \exp \left[-\omega\left(k_{2m-1}\right)t \right] \nonumber\\
& \times \left(-1\right)^{m+1}\cot\left(\frac{k_{2m-1}}{2}\right), \label{1DFPT}
\end{align}
where $\omega(\psi)$ is defined above and $k_{n}= \pi n / {\mathscr L}_y$. This FPT \textcolor{black}{probability density} has multiple time scales. At long times, the FPT \textcolor{black}{probability density} decays exponentially, $F(t) \sim \exp \left(-t/ \hat{t} \right)$, with the TSDT \textcolor{black}{for ${{\mathscr L}_y}\gg1$} given by:
\begin{align}
\hat{t} \approx \frac{{\mathscr L}^{2}_{y}}{\pi^{2} D_0}. \label{1DTSDT}
\end{align}
The MFPT of the distribution in Eq.~(\ref{1DFPT}) is:
\begin{align}
\left\langle t\right\rangle=\frac{2}{{\mathscr L}_y}\sum^{{\mathscr L}_y/2}_{m=1}\left(-1\right)^{m+1}\frac{\cot\left(\frac{\pi(2m-1)}{2{\mathscr L}_y}\right)}{\omega\left(k_{2m-1}\right)} =\frac{{\mathscr L}_y^{2}}{8D_0}.
\end{align}
We also obtain from Eq.~(\ref{1DFPT}) the MPFPT of the distribution:
\begin{align}
t^{\ast}= \frac{\beta}{\pi^{2} } \frac{{\mathscr L}^{2}_{y}}{ D_0} ,\label{betadef}
\end{align}
with $\beta\approx 0.411$ being the positive solution of the transcendental equation
\begin{align}
\sum^{\infty}_{m=0}\left(-1\right)^{m}\left(2m+1\right)^{3} \exp \left[ -\left(2m+1\right)^{2}\beta \right] =0.
\end{align}
These characteristic time scales all scale with the system size and the diffusion coefficient as ${\mathscr L}^{2}_{y}/D_0$, but exhibit different prefactors: $1/\pi^{2} \approx 0.1$ for the TSDT, $1/8$ for the MFPT, and $\beta/(\pi^{2})\approx0.04$ for the MPFPT. The first two time scales are similar, with the TSDT slightly smaller than the MFPT, while the MPFPT is about one order of magnitude smaller than the TSDT and the MFPT.

Figure~\ref{fig1}C shows the perfect agreement between the analytical expression for the 1D FPT \textcolor{black}{probability density} Eq.~(\ref{1DFPT}) and numerical simulations of 1D channels with ${\mathscr L}_y=100$. At long times, the behavior of the FPT \textcolor{black}{probability density} is fully described by a simple exponential decay. Importantly, when considering our percolation system, we see that for aligned fibers ($S=1$) at high occupation fractions, the FPT \textcolor{black}{probability density} approaches the 1D FPT \textcolor{black}{probability density}. This results from a reduction of dimensionality in the dynamics of the system and the emergence of narrow, elongated channels, and will be the focus of \textcolor{black}{Sec.~\ref{Quasi}}.

\section{\label{Quasi}Quasi-1D searching}

For fiber occupation fractions $\phi_C^{2D} < \phi < 1$ and $S=1$, elongated channels with complex cross-sectional shape emerge, and as $\phi$ approaches $\phi_C^{2D}$ from above, the channel structure becomes fractal. We thus focus on how channel \textcolor{black}{shape} changes as a function of the volume occupation fraction and plot in Fig.~\ref{fig2}A the \textcolor{black}{probability density} $F(N)$ of the number $N$ of lattice sites in the $x-z$ cross-sectional area of the channels for different values of $\phi$. For values of $\phi$ just above $\phi_C^{2D}$, the channel size is distributed with an exponential tail but with a clear shoulder. Hence, channels manifest two characteristic size scales, one for large channels and another one for more compact channels. As $\phi$ increases, the distributions shift toward smaller values of $N$; the distribution becomes narrower, and single-exponentially distributed~\cite{Ding}. In the inset of Fig.~\ref{fig2}A, we plot the average channel size in the cross-sectional area $\langle N\rangle$ as a function of $\phi - \phi_{C}$, together with the known relation from percolation theory $\langle N\rangle \sim \mid \phi - \phi_C^{2D} \mid ^{-\gamma}$, with $\gamma = 43/18$~\cite{Stauffer}. Specifically, we show in Fig.~\ref{fig1}C that systems with fiber occupation fractions of $\phi \geq 0.6$ follow the FPT \textcolor{black}{probability density} of the 1D target-finding process, indicating that for narrow channels with values of $\langle N\rangle \leq 10$, the dynamics are effectively 1D. 

To further understand the structure of the channels, we calculate the channel radius of gyration $R_g$, which is defined from:
\begin{equation}
R_g^{2} = \frac{1}{N} \sum_{i=1}^{N} \left( x_i - \langle x \rangle \right) ^{2} + \left(z_i - \langle z \rangle \right)^{2}, \label{Rg}
\end{equation}
with $x_i$ and $z_i$, being the position of a lattice site within the channel's cross section. We now fix a cross-sectional area of the channel to $N=484$ sites and obtain the distribution of the radius of gyration $F (R_g)$, as shown in Fig.~\ref{fig2}B. The distribution shows that despite the fixed value of $N$, channels with different shapes are obtained. Additionally, the lower inset in Fig.~\ref{fig2}B shows that the radius of gyration scales as $R_g \sim N ^{1/d_f}$, with $d_f=91/48$ the fractal dimension of 2D random percolation~\cite{Stauffer, Havlin}. Finally, in the upper inset of Fig.~\ref{fig2}B we plot the variance of $N$ as a function of $\langle N\rangle$, and observe that it increases following the relation Var($N$) $\sim \langle N\rangle ^{2}$.

\begin{figure*}
\centering
\includegraphics[width=0.7\textwidth]{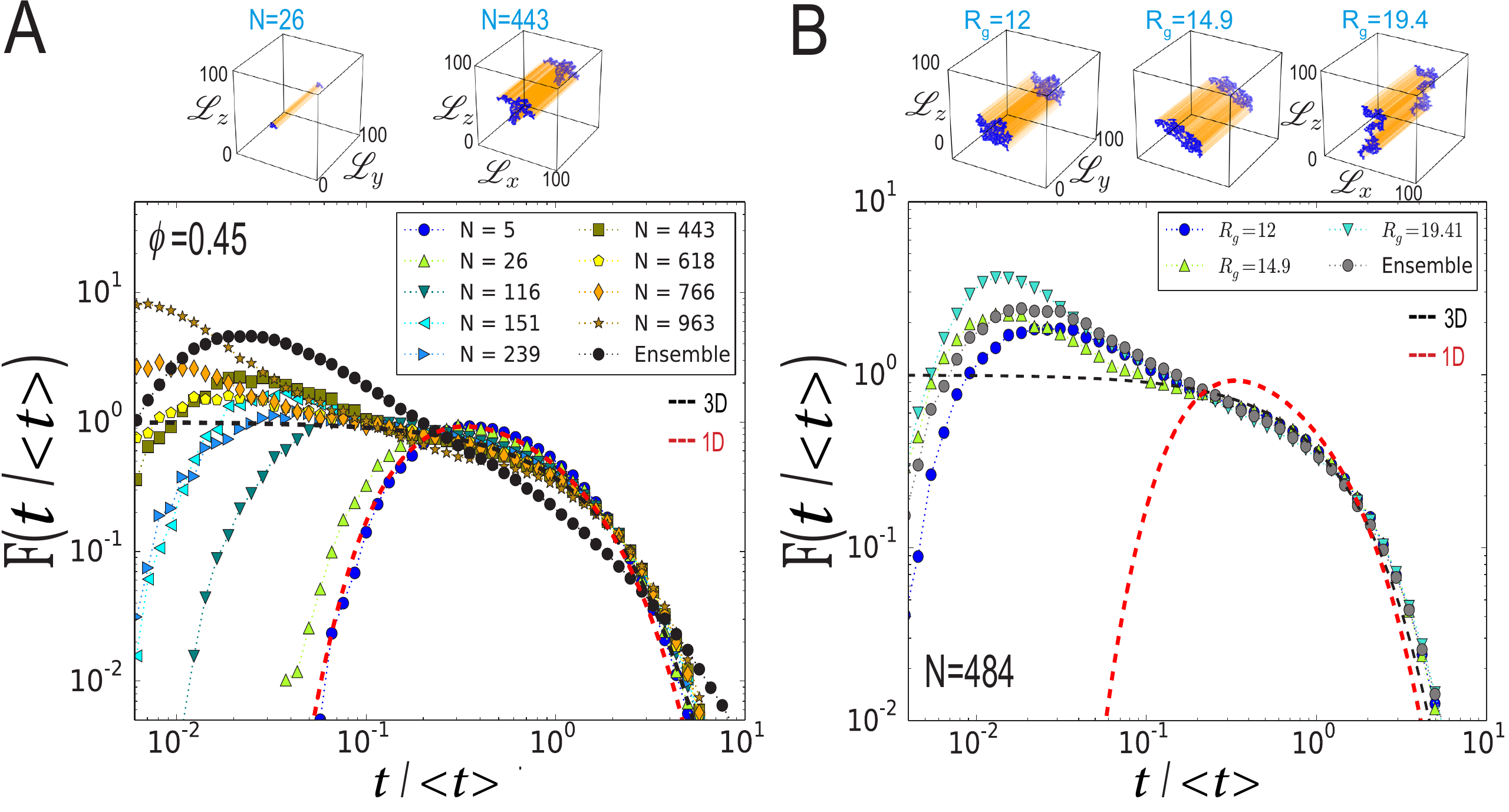} 
\caption{\label{fig:wide}FPT \textcolor{black}{probability densities} for the quasi-1D system. A) FPT \textcolor{black}{probability densities} of nine randomly chosen channels at $\phi=0.45$. B) FPT \textcolor{black}{probability densities} of three channels with different $R_{g}$ but same value of $N=484$. Black dashed lines represent the 3D exponential behavior and the red line is Eq.~(\ref{1DFPT}) for the 1D FPT \textcolor{black}{probability density}. Statistics are performed for $2\times10^{5}$ finding events.}
\label{fig3}
\end{figure*}

\subsection{FPT \textcolor{black}{probability density} in disordered channels}

In Fig.~\ref{fig1}C we show that for channels with ${\mathscr L}_y=100$, $S=1$, and $\phi \gtrsim \phi_C^{2D}$ the FPT \textcolor{black}{probability density} is not the 3D exponential distribution, nor the 1D distribution. Instead, for these cases, as $\phi$ approaches $\phi_C^{2D}$ from above and the channel shapes become fractal, the MPFPT becomes more pronounced and shifts to smaller values. Note that for $\phi =0.42$, the MFPT and the MPFPT differ by almost two orders of magnitude. 

An important point to consider is how the channel size modulates the FPT \textcolor{black}{probability density}. Thus, we next choose a fiber occupation fraction of $\phi=0.45$ and randomly select nine channels with different values of \textcolor{black}{cross-sectional area} $N$ and run our random-walk simulations for each channel configuration, as shown in Fig.~\ref{fig3}A. Here, it becomes evident that the channel's cross-sectional area determines the shape of the FPT \textcolor{black}{probability density}. For example, for $N=5$, the distribution follows the 1D behavior of the situations with high volume occupation fractions. For $N=963$, the distribution has a pronounced maximum at the MPFPT at low FPTs, similar to the case $\phi=0.42,$ in which the fiber density is close to the percolation threshold. The simulations show that as the cross-sectional area of the channel increases, the MPFPT monotonically decreases in values of \textcolor{black}{normalized time $t/\langle t \rangle$}. Similarly, the TSDT monotonically decreases as the cross-sectional area increases. After averaging the contribution of all the channels, the FPT of the ensemble recovers the shape of the distribution for $\phi=0.45$, see Fig.~\ref{fig1}C. The structures of two of the chosen channels are presented in Fig.~\ref{fig3}A; In the case $N=26$, the channel is narrow, and the FPT \textcolor{black}{probability density} is similar to that of the 1D case. The shape of the channel with $N=443$ is more complex, exhibiting internal holes and sharp edges, that support relatively directed trajectories of the tracer toward its target, leading to an FPT \textcolor{black}{probability density} with a pronounced MPFPT at short time scales. 

We now focus on the effect that channel shape has on the FPT \textcolor{black}{probability density} in the natural channels obtained from randomly positioning fibers. For this, we fix the cross-sectional area at $N=484$, and choose channels with three different shapes, with $R_g=12$, 14.9, and 19.4. Figure~\ref{fig3}B shows that, intriguingly, the FPT \textcolor{black}{probability densities} for these three channels seem qualitatively similar, despite the difference in their values of $R_g$. This behavior contrasts with the previously observed effects of $\phi$ and $N$. The FPT \textcolor{black}{probability densities} exhibit a pronounced MPFPT similar to the systems with $\phi \gtrsim \phi_C^{2D}$. Quantitatively, the channel with $R_g = 19.4$ displays a more pronounced MPFPT than the channel with $R_g = 12$. These findings suggest that for the natural channels, the radius of gyration moderately modulates the FPT \textcolor{black}{probability density}. For completeness, we plot in black in Fig.~\ref{fig3}B the FPT \textcolor{black}{probability density} obtained from the whole ensemble of channels with $N=484$. 

\subsection{Synthetic channels}\label{sec:synth}

\begin{figure*}
\centering
\includegraphics[width=0.7\textwidth]{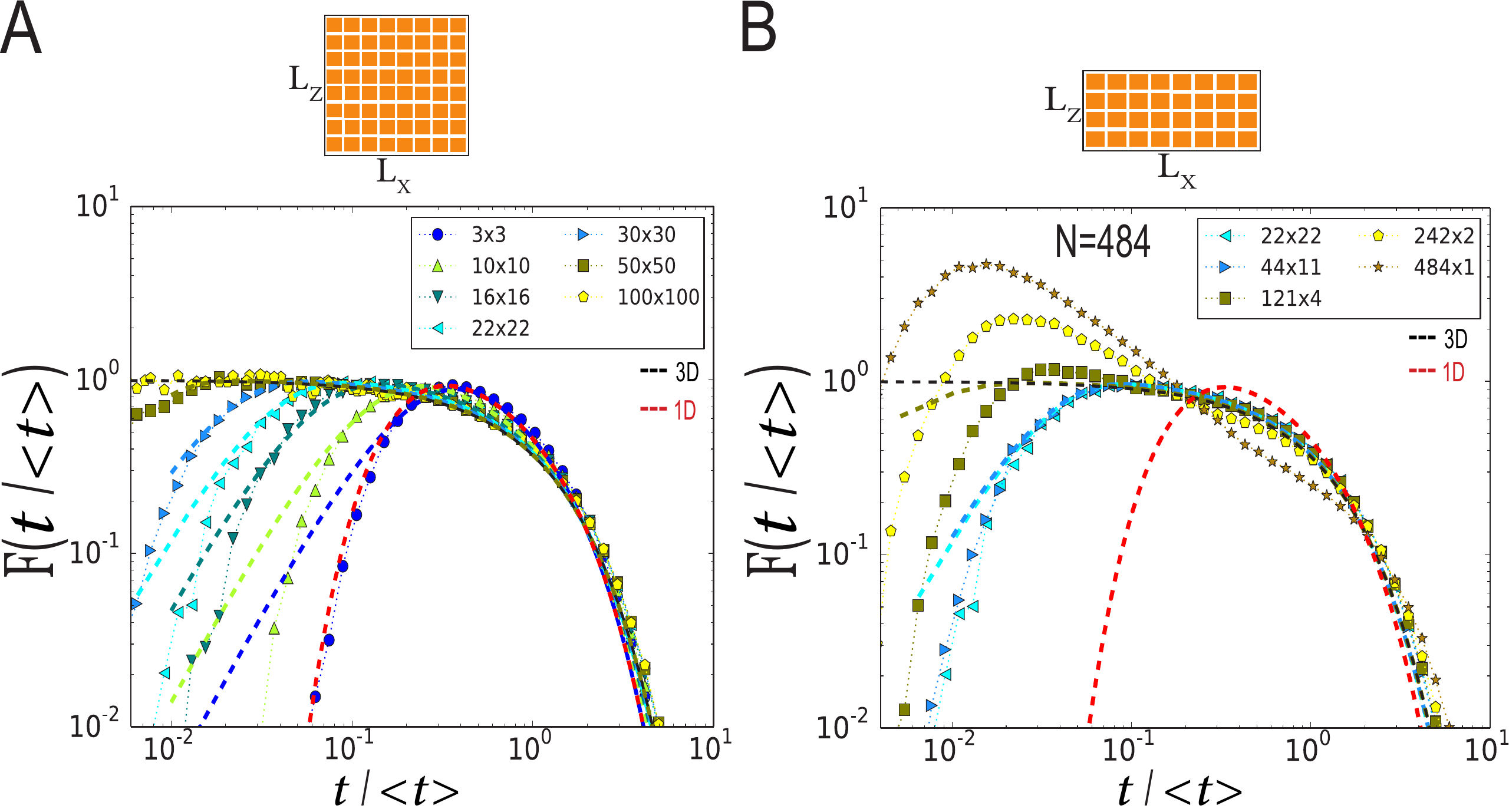} 
\caption{\label{fig:wide}FPT \textcolor{black}{probability densities} for synthetic channels. A) FPT \textcolor{black}{probability densities} for square channels with varying values of $N=L_x \times L_x$. B) FPT \textcolor{black}{probability densities} for rectangular channels with varying aspect ratios for the fixed cross-sectional area $N=484$. We plot in dashed lines the second order approximation, Eq.~(\ref{F2Aprox}) of our solvable model. Black dashed lines represent exponential behavior and the red line is Eq.~(\ref{1DFPT}) for the 1D FPT \textcolor{black}{probability density}. Statistics are performed for $2\times10^{5}$ finding events.}
\label{fig4}
\end{figure*}

To further understand the effect of channel shape and size on the FPT \textcolor{black}{probability density}, we now consider synthetic channels of predefined shapes and sizes, with square cross section $N=L_x \times L_x$ (Fig.~\ref{fig4}A), as well as rectangular channels with different aspect ratios $N=L_x \times L_z$ (Fig.~\ref{fig4}B). \textcolor{black}{This extends what we studied in Sec.~\ref{3DMain} above for cubic domains.} Note that the lengths ${\mathscr L}_i$ are the lattice dimensions in all our simulations, whereas $L_i$ are the dimensions of the synthetic channels considered here. Thus, in general $L_i \leq {\mathscr L}_i $. Specifically, for both channel shapes we choose $L_y = {\mathscr L}_{y}=100$. The radius of gyration for these systems is
\begin{align}
R^{2}_{g}=\frac{L^{2}_{x}+L^{2}_{z}-2}{12} .
\end{align}
These channel geometries allow us to solve the FPT \textcolor{black}{probability density}, based on the evolution equation for the probability density $P\left(\textbf{r},t \right)$ of the particle to be at position $\textbf{r}=(x,y,z)$ at time $t$. For that, we make use of Eqs.~(\ref{fs_derText}) and (\ref{eq1Text}) above, and express $P\left(\textbf{r},t \right)=P_{\textcolor{black}{X}}\left(x,t\right)P_{\textcolor{black}{Y}}\left(y,t\right)P_{\textcolor{black}{Z}}\left(z,t\right)$. We separately obtain $P_{\textcolor{black}{X}}\left(x,t\right)$ and $P_{\textcolor{black}{Z}}\left(z,t\right)$, by implementing reflecting boundary conditions at $x=0,L_x$ and $z=0,L_z$, and $P_{\textcolor{black}{Y}}\left(y,t\right)$ by taking periodic boundary conditions, i.e., $P_{\textcolor{black}{Y}}\left(y,t\right)=P_{\textcolor{black}{Y}}(y+L_y,t)$. Additionally, for simplicity, we assume that $L_y$ is even. After taking the Laplace transform of $P\left(\textbf{r},t \right)$ and using Eq.~(\ref{eq1Text}), we obtain the Laplace transform of the FPT \textcolor{black}{probability density}, or its generating function,
\begin{widetext}
\begin{align}
\tilde{F} \left(\textbf{r}_T,s\right)=\int_0^{\infty} F(r,t) e^{-st} dt  =\frac{\sum\limits^{L_{x}-1}_{n_{x}=0}\sum\limits^{L_{z}-1}_{n_{z}=0}\sum\limits^{L_{y}-1}_{n_{y}=0}\frac{\left(-1\right)^{n_{y}}g\left(n_{x},x_{0},L_{x}\right)g\left(n_{z},z_{0},L_{z}\right)}{s+\omega\left(\frac{\pi n_{x}}{L_{x}}\right)+\omega\left(\frac{\pi n_{z}}{L_{z}}\right)+\omega\left(\frac{2\pi n_{y}}{L_{y}}\right)} }{\sum\limits^{L_{x}-1}_{n_{x}=0}\sum\limits^{L_{z}-1}_{n_{z}=0}\sum\limits^{L_{y}-1}_{n_{y}=0}\frac{g\left(n_{x},x_{0},L_{x}\right)g\left(n_{z},z_{0},L_{z}\right)}{s+\omega\left(\frac{\pi n_{x}}{L_{x}}\right)+\omega\left(\frac{\pi n_{z}}{L_{z}}\right)+\omega\left(\frac{2\pi n_{y}}{L_{y}}\right)} } ,\label{final_fpt2}
\end{align}
\end{widetext}
with
\begin{align}
g\left(n,x,L\right)=\left(2-\delta_{n,0}\right)\cos^{2}\left(\frac{\pi n\left(2x-1\right)}{2L}\right),
\end{align} 
and the function $\omega(\psi)$ defined in Sec.~\ref{3DMain} above. For the complete derivation, see Appendix~\ref{3D}. Similar expressions for the generating function of the FPT \textcolor{black}{probability density} were derived in~\cite{Giuggioli} for a $d$-dimensional system with arbitrary boundary conditions. Here, we concentrate on thoroughly investigating the specific system at hand of rhombic 3D channels.

From the Laplace transform of the FPT \textcolor{black}{probability density}, we obtain the MFPT
\begin{widetext}
\begin{align}
\left\langle t\right\rangle =\sum_{n_{x},n_{y},n_{z}}\frac{\left[1-\left(-1\right)^{n_{y}}\right] 
g\left(n_x,\left\lceil L_x/2 \right\rceil,L_x\right) g\left(n_z,\left\lceil L_z/2 \right\rceil,L_z\right)}
{\Omega_{0}\left(\frac{n_{x}}{L_{x}},\frac{n_{z}}{L_{z}},\frac{n_{y}}{L_{y}}\right)}. \label{mean_fpt_Text}
\end{align} 
\end{widetext}
We note that the sum over $n_{x}$, $n_{y}$, $n_{z}$ in Eq.~(\ref{mean_fpt_Text}) includes all values of $n_{x},n_{y}$ and $n_{z}$ between \textcolor{black}{$0$ and $L_{x}-1,L_{y}-1$ and $L_{z}-1$, respectively, except for the single point $n_{x}=n_{y}=n_{z}=0$}, such that $\Omega_{0}$ is always positive.

In general, inverting the Laplace transform of the FPT \textcolor{black}{probability density} is not trivial. Therefore, we approximate the Laplace transform $\tilde{F}$ of the FPT by functions $\tilde{F}_{\textcolor{black}{M}}$ which agree in the first $\textcolor{black}{M}$ terms in their Taylor expansion. For $M=1,2$ these approximations can be inverted explicitly by
\begin{align}
F_{1}\left(t \right)=\frac{1}{\left\langle t\right\rangle}e^{-t/\left\langle t\right\rangle} , \label{F1Aprox} 
\end{align}
\begin{align}
F_{2}\left(t \right)=\frac{2}{\sqrt{2\left\langle t^{2}\right\rangle-3\left\langle t\right\rangle^{2}}}\exp\left[-\frac{\left\langle t\right\rangle t}{2\left\langle t\right\rangle^{2}-\left\langle t^{2}\right\rangle}\right]\nonumber \\
\times \sinh\left(\frac{\sqrt{2\left\langle t^{2}\right\rangle-3\left\langle t\right\rangle^{2}}}{2\left\langle t\right\rangle^{2}-\left\langle t^{2}\right\rangle}t\right). \label{F2Aprox} 
\end{align}
Higher order approximations are given by
\begin{align}
F_{\textcolor{black}{M}}=\sum^{\textcolor{black}{M}}_{\textcolor{black}{\nu}=1}A_{\textcolor{black}{\nu}}e^{\xi_{\textcolor{black}{\nu}}t} ,
\end{align}
where $\xi_{\textcolor{black}{\nu}}$ are the roots of an $\textcolor{black}{M}$'th order polynomial. The polynomial and the coefficients $A_{\textcolor{black}{\nu}}$ are further detailed in Appendix~\ref{3D}. We note that the second order approximation $F_{2}(t)$ is valid only if
\begin{align}
2\left\langle t\right\rangle^{2}\geq\left\langle t^{2}\right\rangle .
\end{align}
The range of validity of higher order approximations is smaller, as discussed in Appendix~\ref{3D}. In particular, we find that if both $L_{x}$ and $L_{z}$ are smaller than $L_{y}$, then the second order approximation is valid. 

Figure~\ref{fig4}A shows the FPT \textcolor{black}{probability densities} of our simulations for square channels with different \textcolor{black}{cross-sectional areas} $N$. Additionally, we plot in dashed lines in Fig.~\ref{fig4}A our second-order approximation $F_2(t)$ of the FPT \textcolor{black}{probability density}, showing excellent agreement with our simulations \textcolor{black}{except for very short times}. Interestingly, for all considered cases, the shape of the FPT \textcolor{black}{probability densities} differ from the ones that are naturally obtained from percolation. For small squares, the distribution follows the 1D behavior, as shown above for narrow channels. The MPFPT shifts toward lower FPTs when increasing the square lateral size, but without sharply increasing its peak, in contrast to the natural percolation case. Note that the shape of the distribution gradually changes from the 1D limit to the exponential behavior seen for the cubic case with $\phi=0$ and $L_x=L_y=L_z=100$. We conclude that channel size strongly modulates the magnitude of the MPFPT, in some cases making it more than one order of magnitude smaller than the MFPT. However changing channel size using the simplest square-shaped synthetic channels is not enough in order to capture the qualitative evolution of the FPT \textcolor{black}{probability density} seen above for natural channels.

\begin{figure}
\centering
\includegraphics[width=\columnwidth]{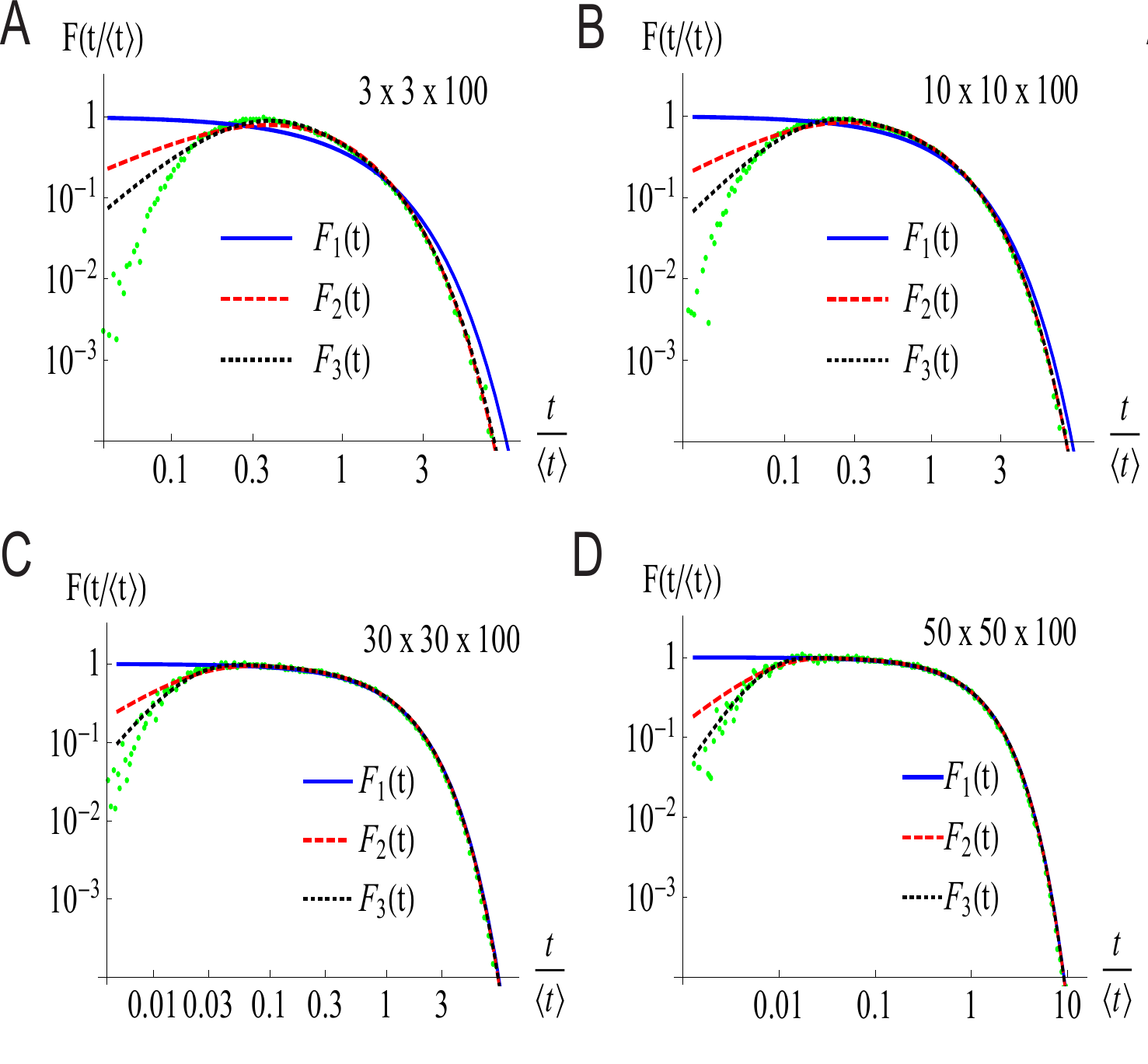} 
\caption{Comparison between the analytical approximation for the FPT \textcolor{black}{probability density}, $F(t)$, and the numerical results (green) for square cross sections. The different colors are the first $F_{1}(t)$ (blue), second $F_{2}(t)$ (red), and third $F_{3}(t)$ (black) order approximations.}
\label{fpt_dist_comp_log}
\end{figure}

In order to better see the location of the MPFPT and the behavior of the FPT at small times, we compare the FPT \textcolor{black}{probability density} from the numerical results to the analytical approximations up to third order. For narrow channels, the second and third order approximations give progressively better results for short times, see Fig.~\ref{fpt_dist_comp_log}. For wider channels, the first order approximation (a simple exponent) agrees very well with the results at times longer than the MPFPT, as shown in Fig.~\ref{fpt_dist_comp_log}C-D. 

\begin{figure}
\centering
\includegraphics[width=\columnwidth]{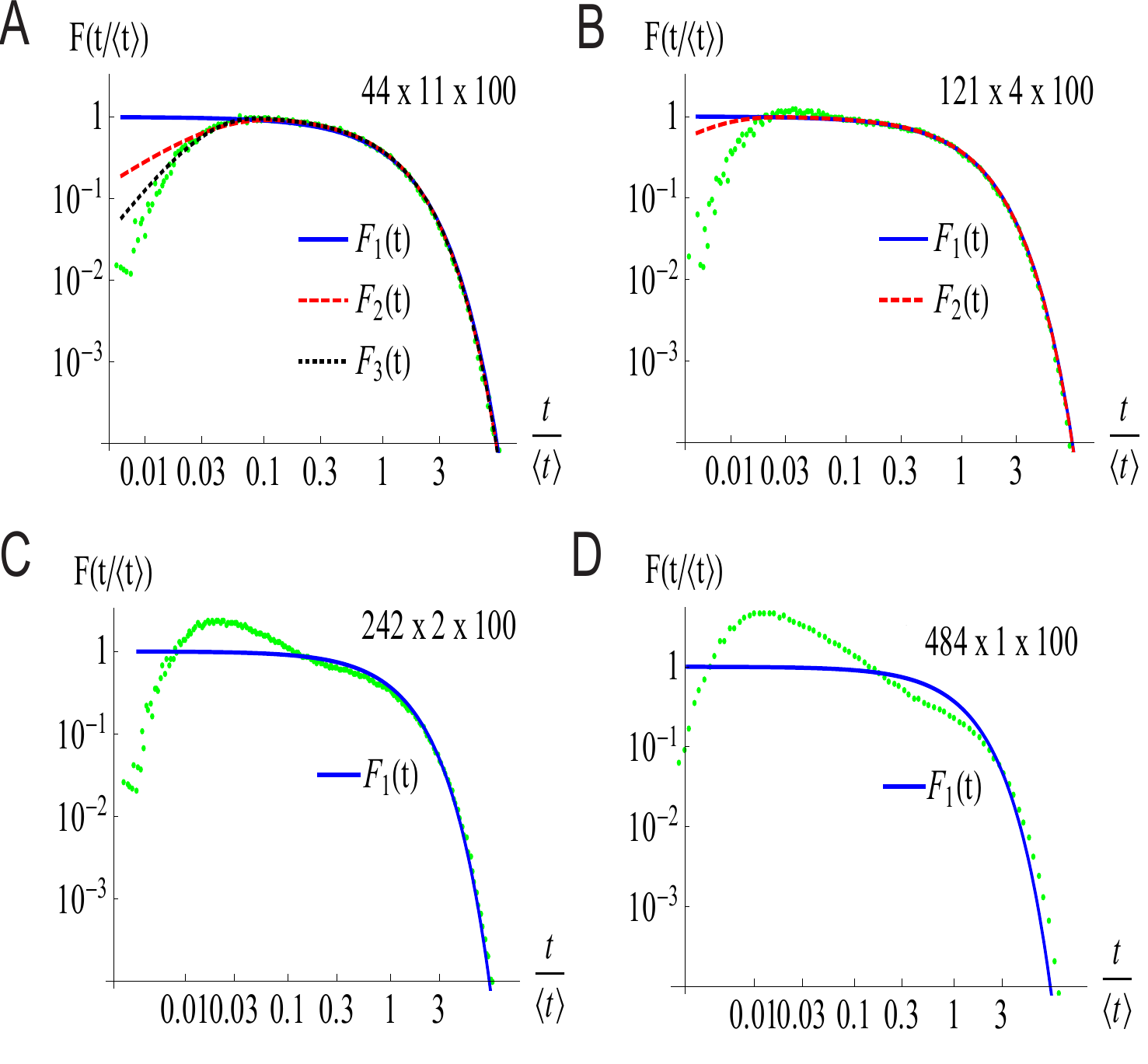} 
\caption{Comparison between the analytical approximation for the FPT \textcolor{black}{probability density}, $F(t)$, and the numerical results (green) for rectangular cross sections of the same area $N=484$. The different colors are the first order $F_{1}(t)$ (blue), second order $F_{2}(t)$ (red), and third order $F_{3}(t)$ (black) approximations. The higher order approximations are not shown for the narrow rectangular cross sections, because they are invalid there.}
\label{fpt_dist_rec}
\end{figure}

\begin{figure}
\centering
\includegraphics[width=\columnwidth]{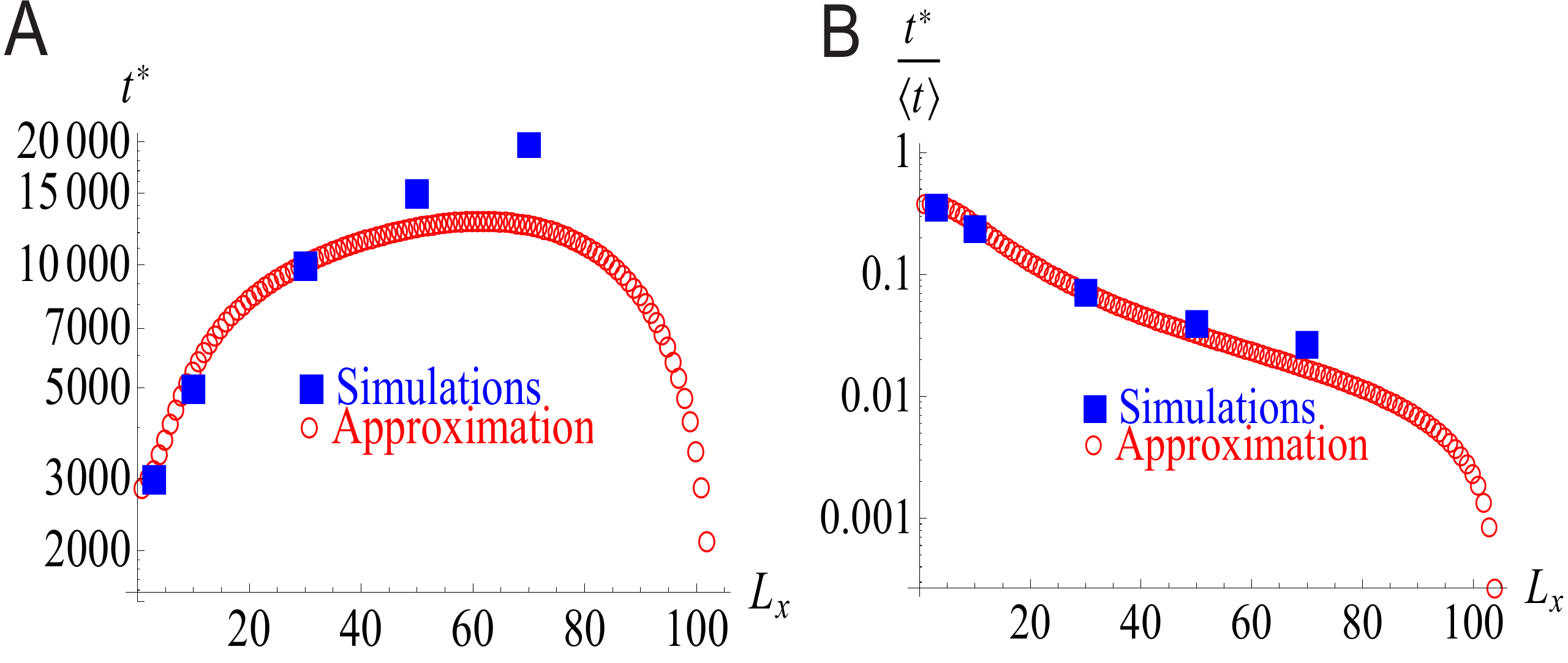} 
\caption{The MPFPT $t^{\ast}$ (A) and the ratio between the MPFPT and the MFPT $t^{\ast}/\left\langle t\right\rangle$ (B) for square cross sections of size $L_{x}\times L_{x}$ as a function of $L_{x}$. In all cases $L_{y}=100$. The empty circles are the results of the approximation, Eq.~(\ref{MPFPT}), and the full squares are the simulation results.}
\label{mpfpt_fig}
\end{figure}

From our solvable model we can also obtain an approximation for the location of the MPFPT, by looking at the maximum of $F_2(t)$:
\begin{align}
t^{\ast}=\frac{ 2\left\langle t\right\rangle^{2}  -  \left\langle t^{2}\right\rangle}{\sqrt{2\left\langle t^{2}\right\rangle-3\left\langle t\right\rangle^{2}}}\cosh^{-1}\left[\frac{\left\langle t\right\rangle}{\sqrt{4\left\langle t\right\rangle^{2}-2\left\langle t^{2}\right\rangle}}\right]. \label{MPFPT}
\end{align}
{\color{black}From Figs.~\ref{fig4} and \ref{fpt_dist_rec} we see that the position of the MFPT is accurately captured, however from Figs.~\ref{fig4}B and \ref{fpt_dist_rec}B we find that the magnitude of the peak is not.} Figure~\ref{mpfpt_fig} shows the very good agreement between the approximation for the MPFPT, Eq.~(\ref{MPFPT}), and the simulation results. We find that for square cross sections the MPFPT has a maximum value and that it vanishes for $L_{x}$ close to the system's length $L_{y}$, while the ratio between the MPFPT and the MFPT is a decreasing function of the cross-section size. Note that the location of the peak at the MPFPT as predicted by the second order approximation agrees very well with the numerical results, and there is no appreciable improvement given by the third order approximation in this regard, even for the narrower channels shown in Fig.~\ref{fpt_dist_comp_log}A-B.

\begin{figure*}
\centering
\includegraphics[width=0.9\textwidth]{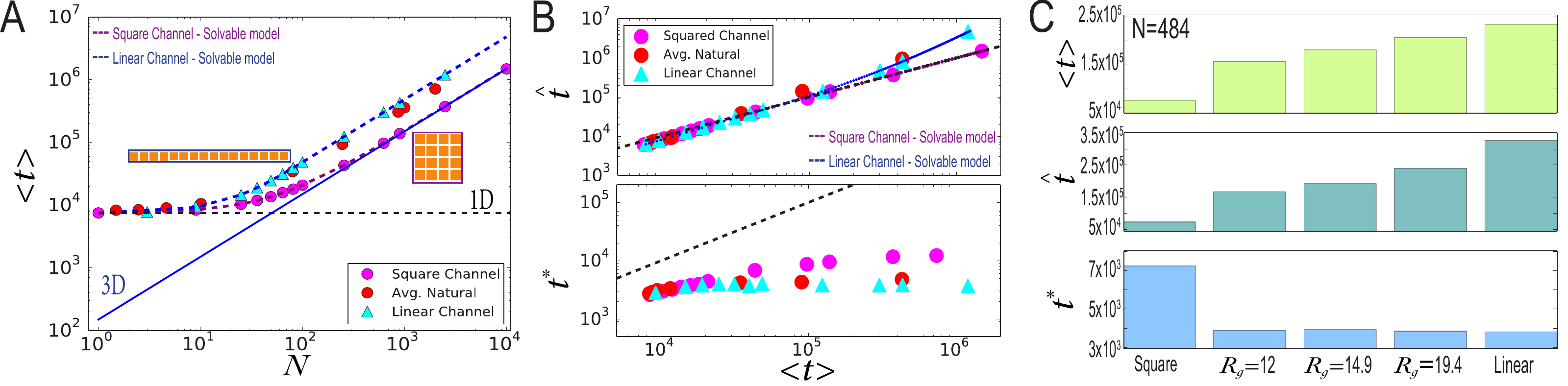} 
\caption{\label{fig:wide}The characteristic time scales are modulated by channel size and shape. A) The MFPT as a function of $N$ for square and linear channels, and channels from the natural percolation model. Purple and blue dashed lines represent Eq.~(\ref{mean_fpt_Text}) for square and linear channels, respectively. B) TSDT (upper panel) and MPFPT (lower panel) as a function of the MFPT for different channel shapes. Black dashed lines are the identity line of slope one. C) Modulation of the MFPT, TSDT, and MPFPT for different channel structures for a fixed value of $N=484$. Statistics are performed for $2 \times 10^{5}$ finding events.}
\label{fig5}
\end{figure*}

Next, we fix the cross-sectional area at $N=484$ and consider rectangular channels with different aspect ratios, see Figs.~\ref{fig4}B and \ref{fpt_dist_rec}. The FPT in the rectangular case $44 \times 11$ is distributed similarly to the square shape $22 \times 22$. As the aspect ratio increases \textcolor{black}{in the $121 \times 4$ system}, the FPT \textcolor{black}{probability density} exhibits a pronounced peak around the MPFPT, in clear contrast to the shape of the FPT \textcolor{black}{probability densities} of square channels. Larger values of the aspect ratio further increase the peak of MPFPT and shift its value toward lower values of FPT/MFPT. For the fully elongated channel $484 \times 1$, with a width of one lattice site, the MPFPT is almost two orders of magnitude smaller than the MFPT, stressing the strong effect of the channel shape on the FPT \textcolor{black}{probability density}. We note that our solvable model does not entirely capture the complex shape of FPT \textcolor{black}{probability densities} for elongated channels, as shown in Fig.~\ref{fpt_dist_rec}.

\subsection{Channel structure modulates the characteristic time scales}

We now examine the effect of channel size and shape on the three different time scales MFPT, MPFPT, and TSDT. First, we plot in Fig.~\ref{fig5}A the MFPT for different channel geometries as a function of $N$. We observe that for small cross-sectional areas of $N < 10$ ($10\%$ of the channel's length ${\mathscr L}_{y}=100$), the MFPT effectively follows a 1D dynamics. In this regime, channels with a square or elongated shape or obtained from our percolation model, exhibit the same MFPT. Here, the MFPT scales with the source-to-target distance squared, ${\mathscr L}_y^{2}$. As $N$ increases, the 3D shape of the channels starts to affect the dynamics of the tracer, and the MFPT rapidly increases in all considered channel shapes. Therefore, modulation of the ratio between $N$ and ${\mathscr L}_{y}$ controls the reduction of dimensionality in the target-finding dynamics. Remarkably, for values of $N \geq 10$, channel shape influences the MFPT. Linear channels with a width of one lattice site, exhibit the highest MFPT. On the contrary, for a given value of $N$, square channels have the lowest MFPT. Note that for values of $N> 200$, square channels behave as the 3D system characterized by Eq.~(\ref{mfpt_asymTex}). Our solvable model captures the MFPT for these two extreme channel shapes, as shown by the blue and purple dashed lines. Interestingly, the ensemble of natural channels quantitatively behaves more similarly to the synthetic linear channels than to the square channels. We note that for this case, we use the average channel size obtained for each value of $\phi$, instead of a fixed value of $N$. The MFPTs from these channels are located between the square and linear channels. 

Motivated by the MFPT dependence on channel structure, we study how the MPFPT and the TSDT correlate with the MFPT. In general, we see that for all channel shapes, the TSDT is highly correlated with the MFPT, see Fig.~\ref{fig5}B upper panel, and therefore, with the cross-sectional area $N$ of the channel, see Appendix~\ref{TSDT}. Moreover, we see that the TSDT-MFPT correlation is affected by the channel shape. Specifically, we see that for square channels, the MFPT converges toward the TSDT, increasing with $N$. Instead, for linear channels with $N \geq 10$, the TSDT becomes larger than the MFPT and rapidly grows with $N$. Note the excellent agreement of our solvable model with our simulations. For channels originated from our percolation model, the TSDT follows a similar behavior to that of the linear channels.

The short-time behavior of the FPT \textcolor{black}{probability density}, which is characterized by the MPFPT, is also channel shape-dependent. We plot in the lower panel of Fig.~\ref{fig5}B the MPFPT as a function of the MFPT. In general, the MPFPT corresponds to events where the tracer finds its target in a relatively direct manner. We consider square channels with values of $N \leq 70^{2}$ and see that the MPFPT increases with the MFPT. As the channel's cross-sectional area increases, the trajectories become less directed, and the MPFPT increases. Importantly, we showed in Fig.~\ref{mpfpt_fig} \textcolor{black}{above} that the MPFPT sharply decreases as the system dimensions $L_x$ and $L_z$ approach ${\mathscr L}_{y}$ and the lattice becomes cubic. In this limit, the MPFPT corresponds to perfectly directed trajectories toward the target. Still, the probability of such events is very low, and the particle needs to scan, on average, the systems volume to find its target. For linear channels, the MPFPT starts increasing for low values the MFPT but then saturates and remains constant as the MFPT increases. Here, the narrow structure of the channel ensures that the directed trajectories toward the target are similar, despite the differences in the channel length. In case the tracer leaves the vicinity of the target and diffuses toward the edges of the channel, the MFPT increases, and in particular, the TSDT happens to dominate the long-time behavior of the FPT \textcolor{black}{probability density}. Similar to the behavior of the MFPT and TSDT, the MPFPT for channels from our percolation model is quantitatively similar to the linear channels, indicating that the local fractal structure of the channels confines the directed trajectories toward the target. 

We now fix a value of $N=484$ and quantify how the channel structure, characterized through $R_g$, modulates the different time scales. We take channels with varying values of $R_g=12$, 14.9, and 19.4, from the natural percolation model and compare their time scales with the square and linear channels. We show in the upper panel of Fig.~\ref{fig5}C that the MFPT increases with $R_g$, i.e., channel elongation. As $R_g$ increases, the time needed for the tracer to come back to the vicinity of its target increases, spending time in regions of the channel where the target is not present, and thus increasing the MFPT. Remarkably, although all cases have the same channel volume $V=N\times {\mathscr L}_{y}$, channel shape modulates the MFPT, implying that the MFPT no longer defines the diffusion-controlled process, affecting the typical notion of reaction dynamics. Similarly, the TSDT increases with the channel elongation, however the MPFPT is highest for square channels. Specifically, the MPFPT is similar for the three considered natural channels and the linear channel. These results support the notion of local channel compactness supporting the fast trajectories toward the target. 

\section{\label{Intermediate}Intermediate fiber Alignment}

\begin{figure}
\centering
\includegraphics[width=\columnwidth]{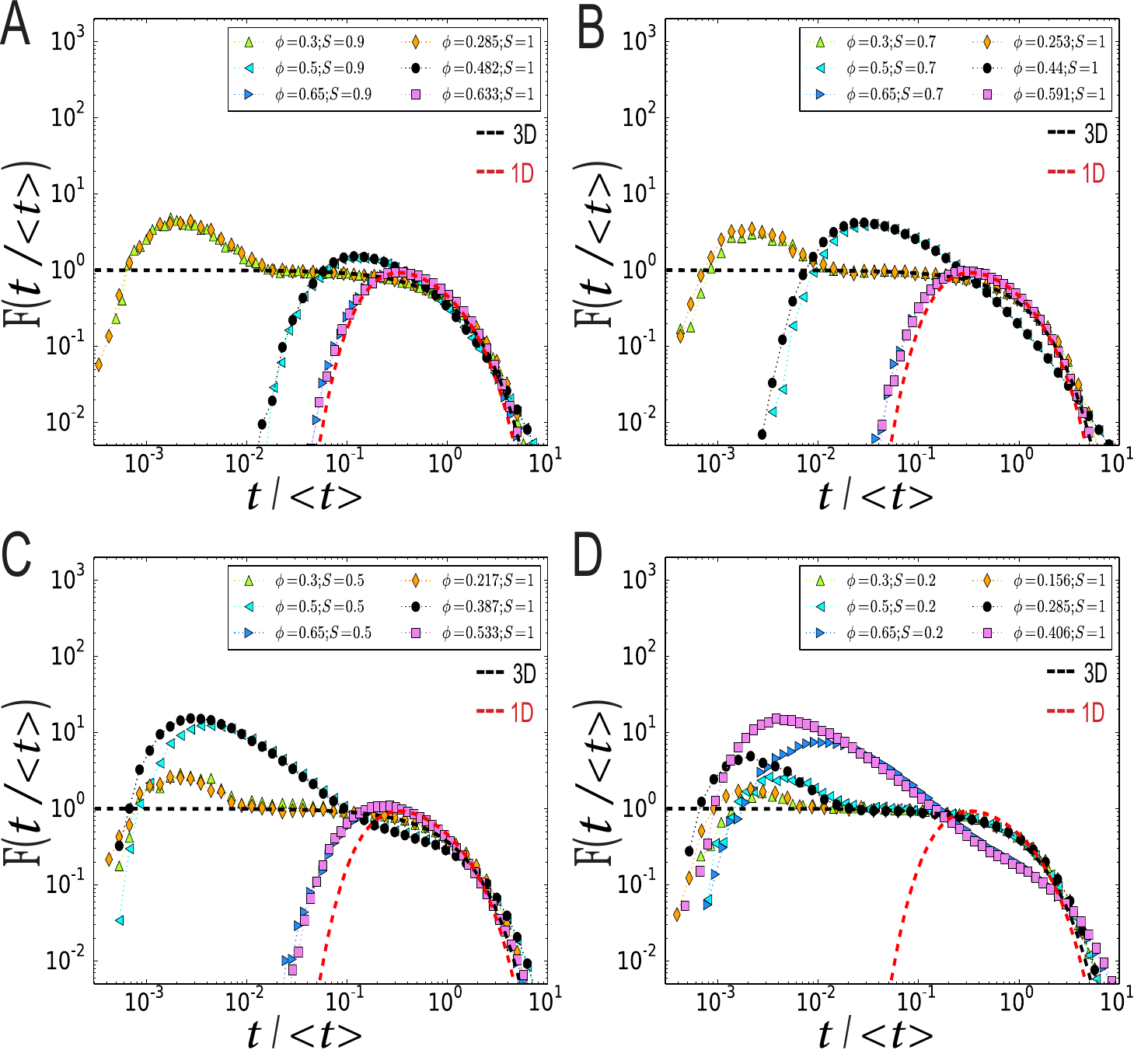} 
\caption{{\color{black}FPT \textcolor{black}{probability densities} for systems with intermediate nematic order parameters $0<S<1$ can be mapped onto systems with aligned fibers $S=1$ via Eq.~(\ref{arealDen}). Black dashed lines represent exponential behavior and the red line is Eq.~(\ref{1DFPT}) for the 1D FPT \textcolor{black}{probability density}. Statistics are performed for $2\times10^{5}$ finding events.}}
\label{fig6}
\end{figure}

{\color{black}Finally, we consider intermediate values of the fiber nematic order parameter $S$ and numerically obtain the FPT \textcolor{black}{probability densities} for different values of $\phi$, as shown in Fig.~\ref{fig6}. We note that all the systems with $S \neq 1$ undergo a drilling percolation transition at some $S$-dependent critical density $\phi_C^{3D}(S)$. For example, for the cases $S=0.5$, the critical density is $\phi_C^{3D} \approx 0.83$~\cite{Gomez4}. For values of $\phi > \phi_C^{3D}$, the FPT diverges. In general, for all the considered values of $S$, similarly to $S=1$, the FPT \textcolor{black}{probability density} has an exponential tail with deviations at short times. For systems with values of $S \geq 0.5 $ and high values of $\phi \geq 0.65$, the FPT \textcolor{black}{probability density} qualitatively follows the shape of the distribution for the 1D case. Here, elongated channels along the $y$-axis emerge, effectively reducing the dimensionality of the dynamics. On the contrary, for intermediate values of $\phi$ and low values of $S \leq 0.2$, no elongated channels are formed, and the FPT \textcolor{black}{probability density} exhibits a pronounced MPFPT. Remarkably, for increasing fiber densities, the FPT \textcolor{black}{probability densities} for systems with $S \neq 1$ behave qualitatively similar to a case with $S=1$, indicating that we can map the systems with $S \neq 1$ onto the $S=1$ case. This mapping consists of identifying the fraction of fibers that run into the $x-z$ plane in a case with $S \neq 1$, and then, considering a new system of aligned fibers $S=1$ with such areal fiber density. Specifically, we find the areal density $\hat{\phi}=M/({\mathscr L}_x {\mathscr L}_z)$, by using $p_x=p_z=(1-S)/3$ and $p_y=(2S+1)/3$ in the previously given relation for $\phi$, and solving the third-order equation:}
\begin{align}
\hat{\phi}^{3} \left( \frac{2S^{3} - 3S^{2} +1}{27}     \right) - \hat{\phi}^{2} \left(     \frac{1-S^{2}}{3}      \right)  +  \hat{\phi}  - \phi = 0. \label{arealDen}
\end{align}

\begin{figure}
\centering
\includegraphics[width=\columnwidth]{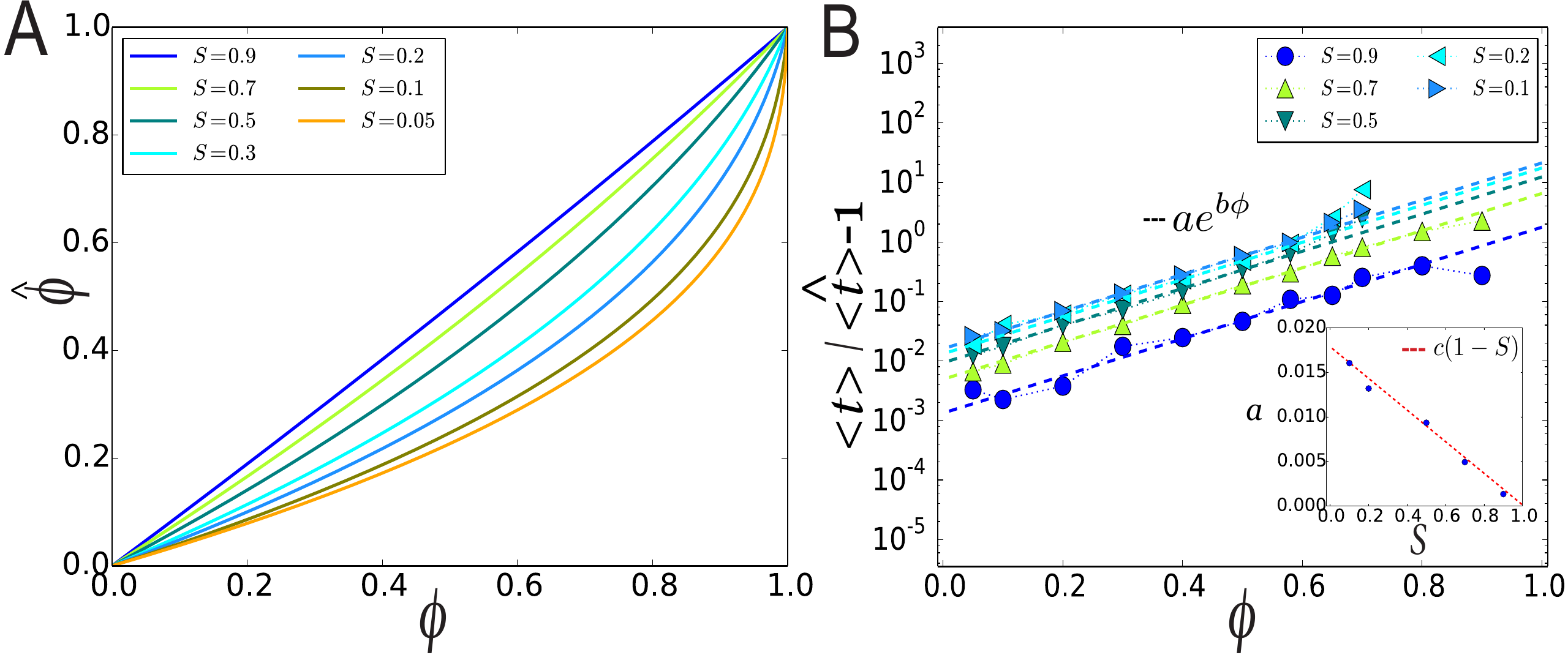} 
\caption{{\color{black}A) Fiber density mapping for systems with $0<S<1$ to systems with $S=1$, Eq.~(\ref{arealDen}), for different values of $S$. B) Relative deviations of the MFPT of systems with $0<S<1$ from the mapped system with $S=1$, for different values of $S$. The relative deviations are fitted to $ae^{b\phi}$, with $b=7.18$. Inset: linear dependence on $S$ of the prefactor $a$. Data is fitted to $(1-S)c$, with $c=0.017$.}}
\label{fig7}
\end{figure}

{\color{black}The areal fiber density on the $x-z$ plane is the real solution of Eq.~(\ref{arealDen}) times $p_y$, i.e., $ \hat{\phi} (2S+1)/3 $, and is presented in Fig.~\ref{fig7}. For example, for $S=0.5$ and $\phi=0.5$, the corresponding mapping is to a system with $S=1$ and $\phi=0.387$. Figure~\ref{fig6} shows the impressive agreement of the FPT \textcolor{black}{probability densities} for the considered mapped systems. Additionally, we plot in Fig.~\ref{fig7}B the relative deviations of the MFPT $\left \langle t \right \rangle$ for systems with $S \neq 1$, from the MFPT $\left\langle \hat{t} \right \rangle$ obtained for the corresponding mapped systems with $S=1$. In general, deviations are small for low values of $\phi$ and high values of $S$. As the systems become more isotropic, and the fiber density increases, the deviations increase as well. Interestingly, the relative deviation $\langle t \rangle / \left\langle \hat{t} \right\rangle - 1$ increases exponentially with fiber density as $a e^{b\phi}$, with $b\approx7.18$ and the prefactor decreasing linearly with S, $a=c(1-S)$ with $c\approx0.017$, as shown by the inset in  Fig.~\ref{fig7}B. Therefore, our model can be applied to relevant biological systems, which in general have fibers oriented with arbitrary nematic order parameters $0 < S < 1$, and not only the two extreme cases $S=0$ and $S=1$.}

\section{\label{Discussion}Discussion}

We demonstrated how the FPT \textcolor{black}{probability density} of a target-finding process is affected by the size and shape of the domain in which the process takes place. In our lattice model, before the system reaches percolation, the FPT process is characterized by an exponential distribution, and the MFPT scales with the available volume. As $\phi$ approaches the critical densities of the system $\phi_C^{3D}$ and $\phi_C^{2D}$, for $S=0$ and $S=1$, respectively, we observed significant deviations from exponential behavior. In the presence of \textcolor{black}{isotropically positioned} fibers, once the system reaches percolation at $\phi_C^{3D}$, the FPT diverges. On the contrary, when fibers are fully aligned, the dynamics are richer, and FPT \textcolor{black}{probability densities} with different shapes emerge. Thus, fiber alignment and fiber density are essential in the understanding of the target-finding process. Additionally, we showed that for intermediate values of fiber densities $\phi \gtrsim \phi_C^{2D}$ of aligned fibers, complex FPT \textcolor{black}{probability densities} are obtained. These distributions are characterized by three time scales: the MPFPT, the MFPT, and the TSTD. We saw that by modulating the channel size and its shape, these time scales were strongly affected. For small cross-sectional areas of $N \leq 10$, channel shape did not affect the FPT \textcolor{black}{probability density} much, and the characteristic time scales remained invariant. In contrast, for channels with larger cross-sectional areas, the FPT \textcolor{black}{probability density} changed with channel shape. Specifically, we saw that linear channels exhibited higher MFPTs than the compact square channel shapes. We also showed that the long-time behavior of the FPT \textcolor{black}{probability density}, characterized by the TSTD, is correlated with the MFPT and less sensitive to channel shape. Moreover, we numerically demonstrated that the short-time behavior of the FPT \textcolor{black}{probability density} is very sensitive to channel shape. We saw that for linear channels, the MPFPT effectively remains constant as the MFPT increases. These results indicate that the confined structure of the linear channels supports relatively directed trajectories of the tracer toward its target. Such directed trajectories remain unaffected as the channel aspect ratio increases since, in those finding events, the tracer does not escape from the vicinity of its target, and the edges of the channels are not explored. Contrarily, for square channels, the MPFPT increases with the MFPT for channels with cross-sectional areas of $N \leq 70$. Here, as $N$ increases along the $x-z$ plane, the directed trajectories defocus, leading to larger MPFPT and MFPT. When considering the ensemble of natural channels from our percolation model, we observed that the FPT \textcolor{black}{probability densities} and its characteristic time scales qualitatively behave more similarly to elongated channels. These observations indicate that the complex fractal shape of the channels support directed trajectories to the target, in a similar way as the linear channels do. \textcolor{black}{We note that the lattice model we use aims to create a clear conceptual understanding of channel emergence through percolation in the lattice and the resulting FPT dynamics on those geometries. The discrete nature of the model imposes limitations on the fiber geometries, in their positions and orientations. It would be interesting to study this complex fibrous system in an off-lattice model.} 

In general, biological systems have a nematic order parameter that lies between the extreme values of $S=0$ and $S=1$. Specifically, the analysis of collagen fiber alignment from ECM porcine urinary bladder obtained an average fiber alignment of $S \approx 0.54$~\cite{Gilbert}. Also, in artificial ECM environments, such as collagen or fibrin hydrogels, the band area between communicating cells reaches values of around $S=0.5$ to 0.7~\cite{Gomez4}. We, therefore, studied the intermediate cases with fiber alignments $0 < S <1$ and showed that they are accurately captured via a mapping onto the case $S=1$. The mapping was achieved by obtaining the occupation fraction of the fibers that are aligned along the preferred direction, and then setting a new system with only those fibers, in such a way that the nematic order parameter is $S=1$. \textcolor{black}{We showed that bellow percolation, the $0 < S <1$ cases exhibit very similar FPT probability densities and characteristic time scales as the mapped case. Thus, the understanding gained for the aligned case ($S=1$) is applicable for more realistic systems with intermediate values of the nematic order parameter $S$. It would be interesting to further study this mapping and its implications.} 

Our findings show that the classical description of diffusion control, given only in terms of the MFPT, is not accurate for biochemical reactions in environments with complex shapes and at a low number of molecules. Instead, the MPFPT and the TSDT should also be considered. We observed that the shape of narrow channels modifies the FPT \textcolor{black}{probability density}, modulating the magnitude of the MPFPT and affecting the MFPT. Interestingly, the MFPT is different for channels with the same volume but different shapes. Consequently, the law of mass action, which states that the reaction rates are directly proportional to each of the reactant concentrations~\cite{Waage}, does not apply to our case with channels that support pronounced MPFPTs. Therefore, for sensory systems in cells that respond to low molecular concentration, two first-passage events will be characterized by very different reaction times~\cite{Grebenkov, Berg2}. A natural extension of our work is to consider the impact of attractive non-specific interactions between the tracer molecule and the elongated obstacles on the characteristic time scales of the target finding process~\cite{Ghosh}.

Previous studies have shown that cells continuously remodel the ECM structure by applying forces to the fiber elements~\cite{Kim, Trubelja, Notbohm1}, and by degrading or generating new ECM fibers~\cite{Kielty, Page-McCaw}. As this remodeling takes place, transport of molecules can be affected, leading to the possibility of biochemical-mechanical signaling feedback~\cite{Gomez4, Jung}. {\color{black} Also, in this work we considered ECM remodeling events generated by cellular activities which are much slower (in the order of minutes to hours~\cite{Natan}) than the transport of molecules. In that case, the picture of static channels in which the molecule travels is appropriate. However, ECM semi-flexible fibers can thermally fluctuate on a time scale comparable to the diffusion of macromolecules~\cite{Lanoiselee, Jahnel}. Thus, another interesting situation to consider in future studies is the dynamic changes of the ECM fibers and the channels they form, for example by modeling fibers that change their position as time progresses.} Our work provides a theoretical basis for such experiments, with a deeper understanding of how fiber remodeling impact molecular transport. Moreover, due to the complex structure of the ECM in bacterial biofilms, our work can provide further understanding of quorum sensing mechanisms and signal transduction in bacterial populations~\cite{Suel}.

\begin{acknowledgments}
We thank Eli Ben-Naim, Gregory Bolshak, Ralf Metzler, and Erdal C. O\u{g}uz for helpful discussions. This work was partially funded by the Tel Aviv University postdoctoral program (D.G), the US-Israel Binational Science Foundation (Y.S), the Israel Science Foundation Grant Numbers 968/16 (Y.S.) and 1474/16 (A.L), the Israel Science Foundation-Israeli Centers for Research Excellence Grant Number 1902/12 (A.L), the Zimin Institute for Engineering Solutions Advancing Better Lives (A.L.), and the National Science Foundation Grant No. NSF PHY-1748958 (Y.S.). Y.S. thanks the Center for Nonlinear Studies at Los Alamos National Laboratory for its hospitality.
\end{acknowledgments}

\appendix
\renewcommand\thefigure{\thesection \arabic{figure}}  
\setcounter{figure}{0}    

\section{3D synthetic channels}
\label{3D}
\setcounter{figure}{0}   

In this appendix we solve the discrete diffusion equation for a particle moving on the cubic lattice in a 3D channel and from it we infer the FPT \textcolor{black}{probability density}. \textcolor{black}{The synthetic channels considered in this appendix are an approximation for the closed natural channels for $S=0$ whose boundaries are formed by the fibers. The synthetic channels do not contain fibers within them.} In Sec.~\ref{sub1} we derive the solution for the diffusion equation. In Sec.~\ref{sub3} we derive the Laplace transform of the FPT \textcolor{black}{probability density}. The moments of the FPT \textcolor{black}{probability density} are derived in Sec.~\ref{sub4}. The approximations of the FPT \textcolor{black}{probability density} are discussed in Sec.~\ref{sec_app_fpt}, and in Sec.~\ref{sub6} we derive the asymptotic expression for the MFPT.

Consider a particle moving on the cubic lattice in a 3D channel. In the $x-z$ plane, the channel has a rectangular cross-section of size $L_{x}\times L_{z}$, with reflecting boundary conditions, while in the $y$ axis it is periodic with length $L_{y}$. We are interested in the distribution of the FPT for the particle to first reach a specific target position $\textbf{r}_{T}=\left(x_{0},\ell,z_{0}\right)$, which differs from its initial location, $\textbf{r}_{0}=\left(x_{0},0,z_{0}\right)$, only in the $y$ coordinate.
In what follows we assume for simplicity that $L_{y}$ is even and that $\ell=L_{y}/2$, such that with the periodic boundary conditions, $\textbf{r}_{0}$ and $\textbf{r}_{T}$ are the farthest away possible along the $y$-axis. We derive the expression for general values of $x_{0}, z_{0}$, but later on concentrate on the specific case where $x_{0}$ and $z_{0}$ are in the middle of the channel.

The FPT \textcolor{black}{probability density}, $F(\textbf{r}_{T},t)$ is related to the probability to find the particle at location $\textbf{r}$ at time $t$ \textcolor{black}{given that it has not yet visited site $\textbf{r}_{T}$}, $P\left(\textbf{r},t\right)$ by~\cite{Siegert1951}
\begin{align}
P\left(\textbf{r}_{T},t\right)=\delta_{\textbf{r}_{T},\textbf{r}_{0}}\delta\left(t\right)+\int^{t}_{0}P\left(\textbf{r}_{0},t-t'\right)F\left(\textbf{r}_{T},t'\right)dt' .\label{fs_der}
\end{align}
The first term on the right hand side of Eq.~(\ref{fs_der}) is the probability that at time $t=0$ the particle is already at location $\textbf{r}_{T}$, and the second term is the probability that \textcolor{black}{before reaching site $\textbf{r}_{T}$ for the first time it was at site $\textbf{r}_{0}$ at time $t-t'$}, and then in the time interval $t'$ it reached site $\textbf{r}_{T}$ once. Taking the Laplace transform of both sides yields
\begin{align}
\tilde{P}\left(\textbf{r}_{T},s\right)=\delta_{\textbf{r}_{T},\textbf{r}_{0}}+\tilde{P}\left(\textbf{r}_{0},s\right)\tilde{F}\left(\textbf{r}_{T},s\right) ,
\end{align}
and therefore, since $\textbf{r}_{T}\neq\textbf{r}_{0}$,
\begin{align}
\tilde{F}\left(\textbf{r}_{T},s\right)=\frac{\tilde{P}\left(\textbf{r}_{T},s\right)}{\tilde{P}\left(\textbf{r}_{0},s\right)} .\label{eq1}
\end{align}

Furthermore, we can decompose the 3D motion of the particle into independent motions along the three axes, such that
\begin{align}
P\left(\textbf{r},t\right)=P_{\textcolor{black}{Y}}\left(y,t\right)P_{\textcolor{black}{X}}\left(x,t\right)P_{\textcolor{black}{Z}}\left(z,t\right) ,
\end{align}
where $P_{\textcolor{black}{J}}\left(j,t\right)$ is the probability for a 1D walker to be at $j$ at time $t$, for $\textcolor{black}{J=X,Y,Z}$. In what follows we derive the 1D probabilities $P_{\textcolor{black}{J}}\left(j,t\right)$ and later transform the probability $P\left(\textbf{r},t\right)$ to Laplace space $\tilde{P}\left(\textbf{r},s\right)$ in order to obtain $\tilde{F}\left(\textbf{r}_{T},s\right)$.

\subsection{Derivation of the 1D probabilities $P_{\textcolor{black}{J}}\left(j,t\right)$}
\label{sub1}

The 1D probabilities $P_{\textcolor{black}{J}}(j,t)$ for $\textcolor{black}{J=X,Y,Z}$ evolve according to the discrete-space diffusion equation,
\begin{align}
3\tau\frac{\partial P_{\textcolor{black}{J}}(j)}{\partial t}=-P_{\textcolor{black}{J}}(j)+\frac{1}{2}\left[P_{\textcolor{black}{J}}(j+1)+P_{\textcolor{black}{J}}(j-1)\right] ,\label{eveq}
\end{align}
where we dropped the explicit dependence of $P_{\textcolor{black}{J}}$ on $t$ for brevity. The prefactor of $3$ on the left hand side comes from the 1D motion accounting for one third of the total 3D motion of the particle, which moves with rate $\tau^{-1}$=1. On a coarse-grained level, the rate $\tau^{-1}$ is related to the diffusion coefficient $D_{0}$ by \textcolor{black}{$D_{0}=\frac{1}{6\tau}$}.
For $\textcolor{black}{J=Y}$ we impose periodic boundary conditions
\begin{align}
P_{\textcolor{black}{Y}}(y)=P_{\textcolor{black}{Y}}(y+L_{y}) ,
\end{align}
while for $\textcolor{black}{J=X,Z}$ we impose reflecting boundary conditions
\begin{align}
&3\frac{\partial P_{\textcolor{black}{J}}(1)}{\partial t}=-P_{\textcolor{black}{J}}(1)+\frac{1}{2}\left[P_{\textcolor{black}{J}}(2)+P_{\textcolor{black}{J}}(1)\right] ,\nonumber\\
&3\frac{\partial P_{\textcolor{black}{J}}(L_{J})}{\partial t}=-P_{\textcolor{black}{J}}(L_{J})+\frac{1}{2}\left[P_{\textcolor{black}{J}}(L_{J})+P_{\textcolor{black}{J}}(L_{J}-1)\right] .\label{bound1d}
\end{align}

Assuming a solution of the form
\begin{align}
P_{\textcolor{black}{J}}(j)=e^{-\omega(k)t}e^{ikj} \label{ansatz3}
\end{align}
yields
\begin{align}
3\omega(k)=1-\cos(k) .
\end{align}
The boundary conditions set restrictions on the allowed values of $k$ \textcolor{black}{as $k_{n}=\frac{\pi n}{L_{J}}$}. Using the initial condition $P_{\textcolor{black}{J}}(j,0)=\delta_{j,j_{0}}$, yields for $\textcolor{black}{J=Y}$
\begin{align}
P_{\textcolor{black}{Y}}\left(r_{y},t\right)=\frac{1}{L_{y}}\sum^{L_{y}-1}_{n=0}e^{-\omega\left(\frac{2\pi n}{L_{y}}\right)t}e^{2i\pi ny/L_{y}} ,
\end{align}
and for $\textcolor{black}{J=X,Z}$
\begin{align}
&P_{\textcolor{black}{J}}\left(j,t\right)=\frac{1}{4L_{J}}\sum^{2L_{J}-1}_{m=0}e^{-\omega\left(\frac{\pi m}{L_{J}}\right)t}   \nonumber\\
&\times \left(e^{-i\pi m(j_{0}-1)/L_{J}}+e^{i\pi mj_{0}/L_{J}}\right) \nonumber\\
&\times \left(e^{i\pi m(j-1)/L_{J}}+e^{-i\pi mj/L_{J}}\right) .\label{pxx}
\end{align}
Setting $j=j_{0}$ in Eq.~(\ref{pxx}) yields
\begin{align}
&P_{\textcolor{black}{J}}\left(j_{0},t\right)=\frac{1}{L_{J}}\sum^{2L_{J}-1}_{m=0}e^{-\omega\left(\frac{\pi m}{L_{J}}\right)t}\cos^{2}\left(\frac{\pi m\left(2j_{0}-1\right)}{2L}\right)=\nonumber\\
&=\textcolor{black}{\frac{1}{L_{J}}\left[1+2\sum^{L_{J}-1}_{m=1}e^{-\omega\left(\frac{\pi m}{L_{J}}\right)t}\cos^{2}\left(\frac{\pi m\left(2j_{0}-1\right)}{2L}\right)\right]} .
\end{align}
Note that $\omega(0)=0$, while for $k>0$ $\omega(k)>0$.

\subsection{FPT \textcolor{black}{probability density} in Laplace space}
\label{sub3}

The Laplace transform of the FPT \textcolor{black}{probability density}, or its generating function is therefore
\begin{widetext}
\begin{align}
\tilde{F}\left(\textbf{r}_{T},s\right)=\frac{\sum^{L_{x}-1}_{n_{x}=0}\sum^{L_{z}-1}_{n_{z}=0}\sum^{L_{y}-1}_{n_{y}=0}\frac{1}{s+\omega\left(\frac{\pi n_{x}}{L_{x}}\right)+\omega\left(\frac{\pi n_{z}}{L_{z}}\right)+\omega\left(\frac{2\pi n_{y}}{L_{y}}\right)}e^{2i\pi n_{y}\ell/L_{y}}\cos^{2}\left(\frac{\pi n_{x}\left(2x_{0}-1\right)}{2L_{x}}\right)\cos^{2}\left(\frac{\pi n_{z}\left(2z_{0}-1\right)}{2L_{z}}\right)}{\sum^{L_{x}-1}_{n_{x}=0}\sum^{L_{z}-1}_{n_{z}=0}\sum^{L_{y}-1}_{n_{y}=0}\frac{1}{s+\omega\left(\frac{\pi n_{x}}{L_{x}}\right)+\omega\left(\frac{\pi n_{z}}{L_{z}}\right)+\omega\left(\frac{2\pi n_{y}}{L_{y}}\right)}\cos^{2}\left(\frac{\pi n_{x}\left(2x_{0}-1\right)}{2L_{x}}\right)\cos^{2}\left(\frac{\pi n_{z}\left(2z_{0}-1\right)}{2L_{z}}\right)} .\label{final_fpt}
\end{align}
Setting $\ell=L_{y}/2$ yields
\begin{align}
\tilde{F}\left(\textbf{r}_{T},s\right)=\frac{\sum^{L_{x}-1}_{n_{x}=0}\sum^{L_{z}-1}_{n_{z}=0}\sum^{L_{y}-1}_{n_{y}=0}\frac{1}{s+\omega\left(\frac{\pi n_{x}}{L_{x}}\right)+\omega\left(\frac{\pi n_{z}}{L_{z}}\right)+\omega\left(\frac{2\pi n_{y}}{L_{y}}\right)}\left(-1\right)^{n_{y}}g\left(n_{x},x_{0},L_{x}\right)g\left(n_{z},z_{0},L_{z}\right)}{\sum^{L_{x}-1}_{n_{x}=0}\sum^{L_{z}-1}_{n_{z}=0}\sum^{L_{y}-1}_{n_{y}=0}\frac{1}{s+\omega\left(\frac{\pi n_{x}}{L_{x}}\right)+\omega\left(\frac{\pi n_{z}}{L_{z}}\right)+\omega\left(\frac{2\pi n_{y}}{L_{y}}\right)}g\left(n_{x},x_{0},L_{x}\right)g\left(n_{z},z_{0},L_{z}\right)} ,\label{final_fpt2}
\end{align}
\end{widetext}
with
\begin{align}
g\left(n,x,L\right)=\left(2-\delta_{n,0}\right)\cos^{2}\left(\frac{\pi n\left(2x-1\right)}{2L}\right) .
\end{align}
We now assume that $x_{0}$ and $z_{0}$ are in the center of the cross-section, such that if $L_{x}$ ($L_{z}$) is even then $x_{0}=L_{x}/2$ ($z_{0}=L_{z}/2$), and if $L_{x}$ ($L_{z}$) is odd then $x_{0}=\left(L_{x}+1\right)/2$ ($z_{0}=\left(L_{z}+1\right)/2$). Under this assumption the functions $g\left(n,x,L\right)$ for even and odd values of $L$ are given by
\begin{align}
&g_{even}\left(n,L\right)=\frac{\left(2-\delta_{n,0}\right)}{2}\left[1+\left(-1\right)^{n}\cos\left(\frac{\pi n}{L}\right)\right] ,\nonumber\\
&g_{odd}\left(n,L\right)=\left(2-\delta_{n,0}\right)\frac{1+\left(-1\right)^{n}}{2} .
\end{align}

\begin{figure}
\centering
\includegraphics[width=\columnwidth]{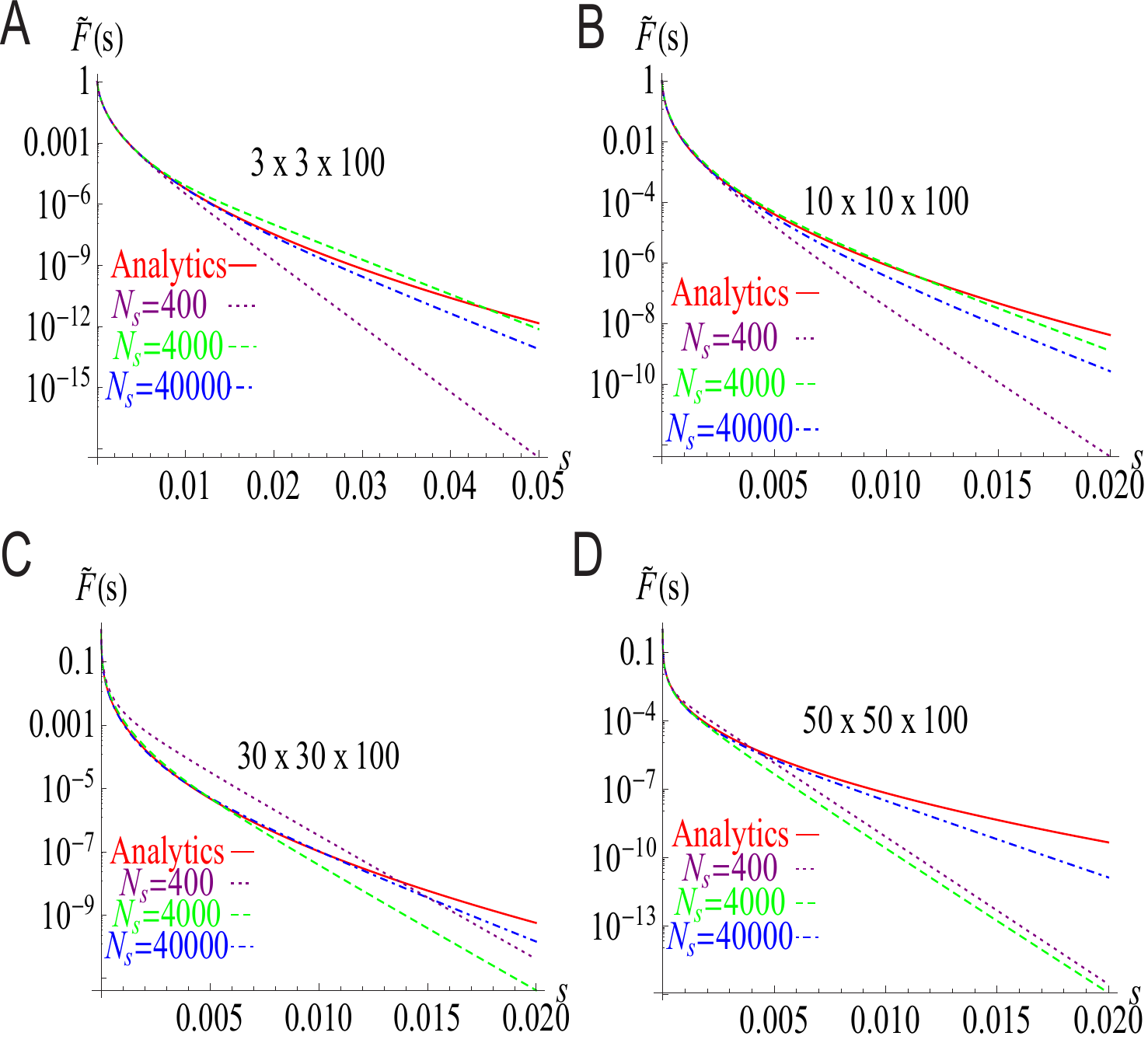} 
\caption{Comparison between the analytical result for the Laplace transform of the FPT, $\tilde{F}(s)$, and the numerical results for different system sizes and different number of simulation runs, $N_{s}$.}
\label{lap_comp}
\end{figure}

We compare in Fig.~\ref{lap_comp} the analytical result for the Laplace transform of the FPT \textcolor{black}{probability density}, $\tilde{F}(s)$, with the numerical results. We find that at small values of $s$ the agreement is excellent, while at higher values of $s$ where the agreement is lacking, the value of $\tilde{F}(s)$ itself is extremely small. The difference is due to limited statistics in the simulations. In order to see that this is indeed the reason, we ran more simulations and saw that the numerical results approach the analytical results as the number of realizations increases.

\subsection{Moments}
\label{sub4}
From the Laplace transform of the FPT \textcolor{black}{probability density} we can find all its moments by 
\begin{align}
\left\langle t^{n}\right\rangle=\left(-1\right)^{n}\left.\frac{\partial^{n}\tilde{F}\left(\textbf{r}_{T},s\right)}{\partial s^{n}}\right|_{s=0} .
\end{align}
The first moment, or MFPT is
\begin{align}
\left\langle t\right\rangle=\sum_{n_{x},n_{y},n_{z}}\frac{\left[1-\left(-1\right)^{n_{y}}\right]g\left(n_{x},L_{x}\right)g\left(n_{z},L_{z}\right)}{\Omega_{0}\left(\frac{n_{x}}{L_{x}},\frac{n_{z}}{L_{z}},\frac{n_{y}}{L_{y}}\right)} ,\label{mean_fpt}  
\end{align}
where 
\begin{align}
\Omega_{0}\left(x,y,z\right)=\omega\left(\pi x\right)+\omega\left(\pi z\right)+\omega\left(2\pi y\right) ,
\end{align}
and the sum over $n_{x},n_{y},n_{z}$ includes all values of $n_{x},n_{y}$ and $n_{z}$ between \textcolor{black}{$0$ and $L_{x}-1,L_{y}-1$ and $L_{z}-1$ respectively, except for the single point $n_{x}=n_{y}=n_{z}=0$}, such that $\Omega_{0}$ is always positive. This special point is excluded since when taking the derivatives of $\tilde{F}(s)$, there are contributions to the sums from both the nominator and denominator of Eq.~(\ref{final_fpt2}), and in this special point $\Omega_{0}=0$ the contributions cancel each other.

The second moment is
\begin{align}
&\left\langle t^{2}\right\rangle=2\sum_{n_{x},n_{y},n_{z}}\frac{\left[1-\left(-1\right)^{n_{y}}\right]g\left(n_{x},L_{x}\right)g\left(n_{z},L_{z}\right)}{\Omega^{2}_{0}\left(\frac{n_{x}}{L_{x}},\frac{n_{z}}{L_{z}},\frac{n_{y}}{L_{y}}\right)} \nonumber\\
&+2\left\langle t\right\rangle\sum_{n_{x},n_{y},n_{z}}\frac{g\left(n_{x},L_{x}\right)g\left(n_{z},L_{z}\right)}{\Omega_{0}\left(\frac{n_{x}}{L_{x}},\frac{n_{z}}{L_{z}},\frac{n_{y}}{L_{y}}\right)} .\label{mean_fpt2}
\end{align}

The third moment is
\begin{align}
&\left\langle t^{3}\right\rangle=6\sum_{n_{x},n_{y},n_{z}}\frac{\left[1-\left(-1\right)^{n_{y}}\right]g\left(n_{x},L_{x}\right)g\left(n_{z},L_{z}\right)}{\Omega^{3}_{0}\left(\frac{n_{x}}{L_{x}},\frac{n_{z}}{L_{z}},\frac{n_{y}}{L_{y}}\right)} \nonumber\\
&+6\left\langle t\right\rangle\sum_{n_{x},n_{y},n_{z}}\frac{g\left(n_{x},L_{x}\right)g\left(n_{z},L_{z}\right)}{\Omega^{2}_{0}\left(\frac{n_{x}}{L_{x}},\frac{n_{z}}{L_{z}},\frac{n_{y}}{L_{y}}\right)}+\nonumber\\
&+3\left\langle t^{2}\right\rangle\sum_{n_{x},n_{y},n_{z}}\frac{g\left(n_{x},L_{x}\right)g\left(n_{z},L_{z}\right)}{\Omega_{0}\left(\frac{n_{x}}{L_{x}},\frac{n_{z}}{L_{z}},\frac{n_{y}}{L_{y}}\right)} .
\end{align}
We note that since $\tilde{F}(s)$ is regular around $s=0$, all the moments exist.

\subsection{Approximations for the FPT \textcolor{black}{probability density}}\label{sec_app_fpt}

Since inverting the full form of the Laplace transform of the FPT \textcolor{black}{probability density} is not practical, we consider approximations that can be inverted. In order to find the $\textcolor{black}{M}$'th order approximation, we first expand $\tilde{F}\left(\textbf{r}_{T},s\right)$ \textcolor{black}{in a Taylor series in $s$}
\begin{align}
\tilde{F}\left(\textbf{r}_{T},s\right)=\sum^{\infty}_{k=0}\frac{\left(-1\right)^{k}\left\langle t^{k}\right\rangle}{k!}s^{k} .\label{ftaylor}
\end{align}
\textcolor{black}{This expansion is valid for $s<\left|\hat{s}\right|$ where $\hat{s}$ is the pole of $\tilde{F}$ with the smallest absolute value. This implies that in the time domain, the approximation is valid for $t>1/\hat{s}$.}
In the next step we \textcolor{black}{further} approximate $\tilde{F}\left(\textbf{r}_{T},s\right)$ by
\begin{align}
\tilde{F}_{\textcolor{black}{M}}\left(\textbf{r}_{T},s\right)=\frac{1}{\sum^{\textcolor{black}{M}}_{k=0}a_{k}s^{k}} ,\label{fapp}
\end{align}
such that when Eq.~(\ref{fapp}) is expanded to $M$'th order in $s$ we retrieve Eq.~(\ref{ftaylor}). \textcolor{black}{We remark that Eq.~(\ref{fapp}) is certainly not the only approximation that yields functions with the same moments as the original function. It was chosen for its simple form.} For $k=0,1,2,3$ we find that
\begin{align}
&a_{0}=1 ,\nonumber\\
&a_{1}=\left\langle t\right\rangle ,\nonumber\\
&a_{2}=\left\langle t\right\rangle^{2}-\frac{1}{2}\left\langle t^{2}\right\rangle ,\nonumber\\
&a_{3}=\left\langle t\right\rangle^{3}-\left\langle t\right\rangle\left\langle t^{2}\right\rangle+\frac{1}{6}\left\langle t^{3}\right\rangle .
\end{align}

Inverting these approximations for the Laplace transform yields successive approximations for $F\left(\textbf{r}_{T},t\right)$. In general, the approximated FPT \textcolor{black}{probability density} is given by
\begin{align}
&F_{\textcolor{black}{M}}\left(\textbf{r}_{T},t\right)=\frac{1}{a_{\textcolor{black}{M}}}\sum^{\textcolor{black}{M}}_{\textcolor{black}{\nu}=1}\frac{\exp\left(\xi_{\textcolor{black}{\nu}}t\right)}{\prod^{\textcolor{black}{M}}_{\textcolor{black}{\nu}'\neq \textcolor{black}{\nu}}\left(\xi_{\textcolor{black}{\nu}}-\xi_{\textcolor{black}{\nu}'}\right)} ,\label{fm_app}
\end{align}
where $\xi_{\textcolor{black}{\nu}}$ are the roots of the polynomial $\sum^{\textcolor{black}{M}}_{k=0}a_{k}\xi^{k}$. For $M=1,2$, it is possible to find an analytical expression for the roots $\xi_{\textcolor{black}{\nu}}$, such that Eq.~(\ref{fm_app}) is explicitly
\begin{align}
&F_{1}\left(\textbf{r}_{T},t\right)=\frac{1}{\left\langle t\right\rangle}e^{-t/\left\langle t\right\rangle} ,\nonumber\\
&F_{2}\left(\textbf{r}_{T},t\right)=\frac{2}{\sqrt{2\left\langle t^{2}\right\rangle-3\left\langle t\right\rangle^{2}}}\exp\left[-\frac{\left\langle t\right\rangle t}{2\left\langle t\right\rangle^{2}-\left\langle t^{2}\right\rangle}\right] \nonumber\\
&\times \sinh\left(\frac{\sqrt{2\left\langle t^{2}\right\rangle-3\left\langle t\right\rangle^{2}}}{2\left\langle t\right\rangle^{2}-\left\langle t^{2}\right\rangle}t\right) .
\end{align}
For $\textcolor{black}{M}\geq3$ we can find the coefficients $a_{k}$ exactly for any system size, and solve the resulting polynomial numerically. 

\begin{figure}
\includegraphics[width=0.6\columnwidth]{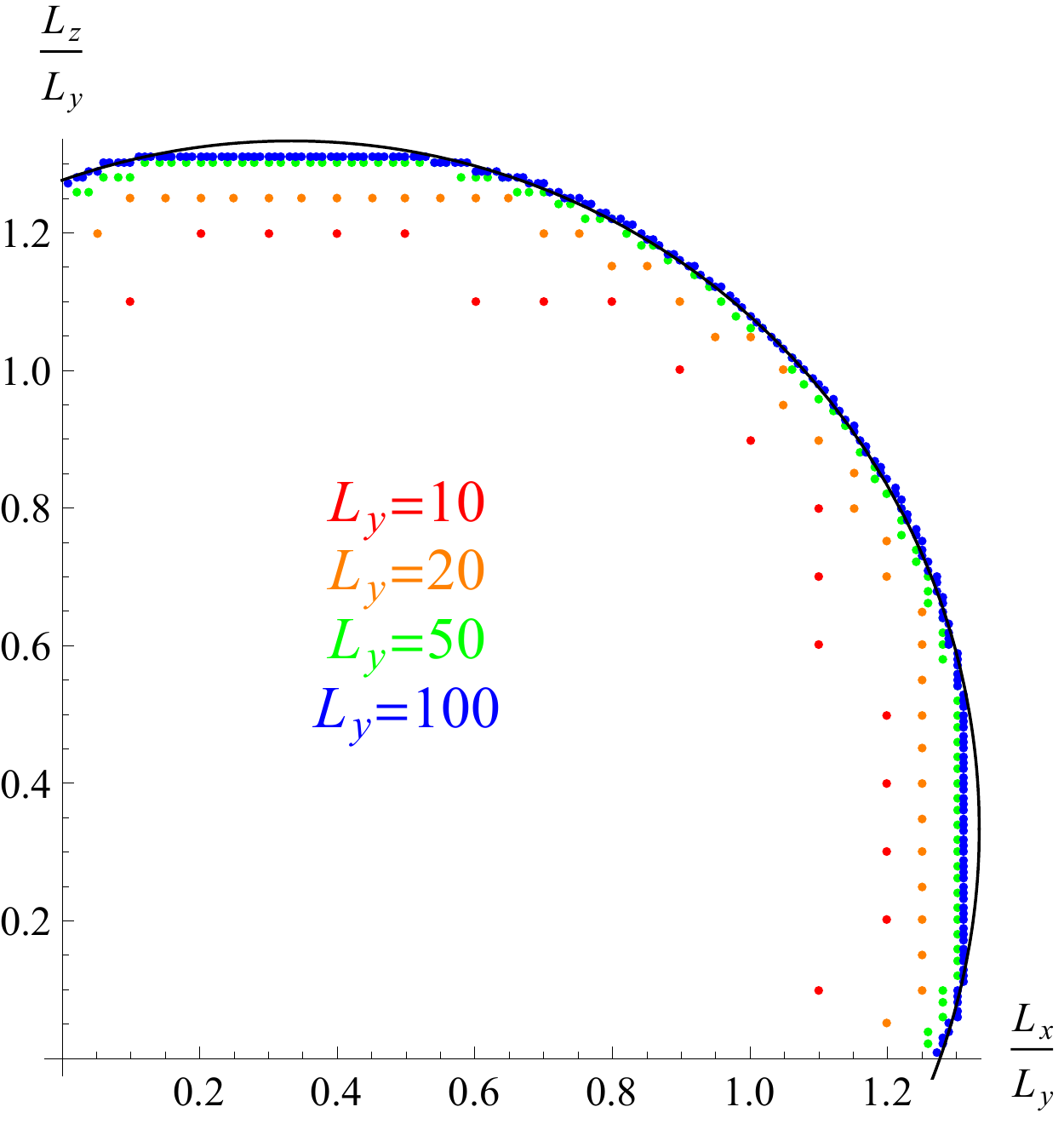}
\caption{The maximum values of $L_{x}/L_{y}$ and $L_{z}/L_{y}$ for which the second order approximation is valid, for $L_{y}=10,20,50$ and $100$. The black line is Eq.~(\ref{valid}) with $\gamma=1/3$.}
\label{phases}
\end{figure}

\begin{figure}
\includegraphics[width=0.7\columnwidth]{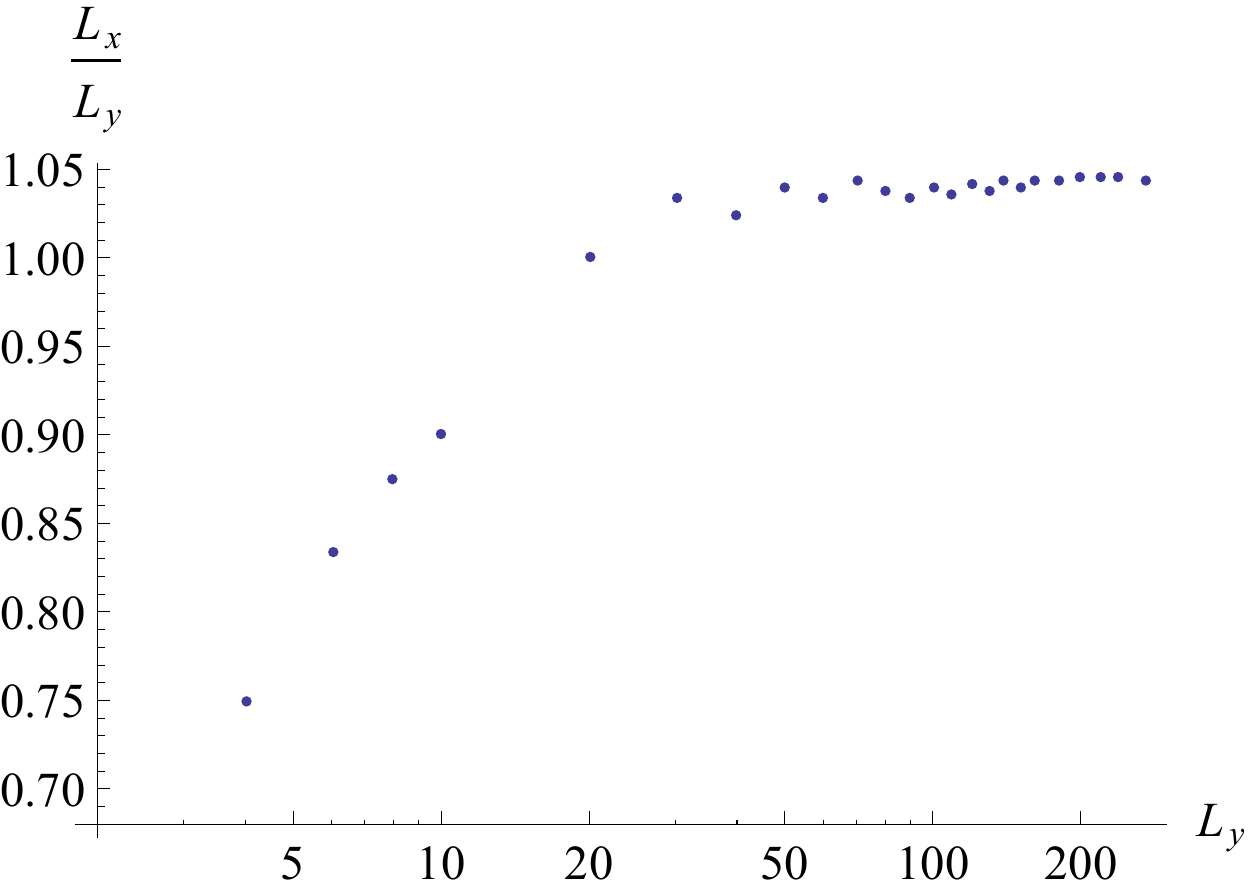}
\caption{The bounds on the value of the parameter $\gamma$ vs. $L_{y}$, as obtained from square cross sections $L_{x}=L_{z}$. The blue squares are obtained from the maximal value of $L_{x}$ for which Eq.~(\ref{valid}) is satisfied.}
\label{gamma}
\end{figure}

Note that these approximations are valid only if the real part of all the roots $\xi_{\textcolor{black}{\nu}}$ is negative. For the first order approximation, $\textcolor{black}{M}=1$, the only root is $-\left\langle t\right\rangle^{-1}$ which is negative. The second order approximation is valid only if
\begin{align}
2\left\langle t\right\rangle^{2}\geq \left\langle t^{2}\right\rangle .
\end{align}
For the higher order approximations, it is straightforward to check that the $M$'th order approximation is valid only if $a_{k}\geq 0$ for all $k\leq \textcolor{black}{M}$. Therefore, the range of validity for each successive approximation is smaller than the previous one. Figure~\ref{phases} shows the values of $L_{x}$ and $L_{z}$ for which the second order approximation is valid for different values of $L_y$. First, we see that the validity depends on the ratios $L_{x}/L_{y}$ and $L_{z}/L_{y}$. We observe numerically that the approximation is valid when
\begin{align}
\left(\frac{L_{x}}{L_{y}}-\gamma\right)^{2}+\left(\frac{L_{z}}{L_{y}}-\gamma\right)^{2}<1 ,\label{valid}
\end{align}
where $\gamma$ depends on $L_{y}$ and is very close to $1/3$. In the case $L_{x}=L_{z}$, Eq.~(\ref{valid}) reduces to $L_{x}/L_{y}<\gamma+1/\sqrt{2}$. In order to evaluate $\gamma$ for larger systems, we concentrate on the case $L_{x}=L_{z}$, and find the largest $L_{x}$ for which the approximation is valid. Figure~\ref{gamma} shows the ratio $L_{x}/L_{y}$ between the largest $L_{x}$ for which the approximation is valid and the system's length $L_{y}$. It appears to converge to a value slightly below $1.05$, i.e. $\gamma$ converges to a value slightly below $0.35$.

In the region where it is valid, the second order approximation, $F_{2}(t)$, has a single maximum at the MPFPT
\begin{align}
t^{\ast}=\frac{2\left\langle t\right\rangle^{2}-\left\langle t^{2}\right\rangle}{\sqrt{2\left\langle t^{2}\right\rangle-3\left\langle t\right\rangle^{2}}}\cosh^{-1}\left[\frac{\left\langle t\right\rangle}{\sqrt{4\left\langle t\right\rangle^{2}-2\left\langle t^{2}\right\rangle}}\right] .\label{tast_app}
\end{align}
Increasing the size of the cross section, $L_{x}\times L_{z}$, or decreasing the length of the channel, $L_{y}$, decreases the value of $t^{\ast}$. Note that although $t^{\ast}=0$ only in the case of an infinite system, its approximation, Eq.~(\ref{tast_app}), reaches $0$ at the edge of validity for the second order approximation when $2\left\langle t\right\rangle^{2}=\left\langle t^{2}\right\rangle$.

\subsection{Asymptotic expressions for large $L_{x},L_{y},L_{z}$}
\label{sub6}
In this section we derive asymptotic expressions for $\left\langle t\right\rangle$ when $L_{x},L_{y}$ and $L_{z}$ are very large. Changing the sums over $n_{x},n_{y}$ and $n_{z}$ in Eq.~(\ref{mean_fpt}) to integrals over $x=n_{x}/L_{x}$, $y=n_{y}/L_{y}$ and $z=n_{z}/L_{z}$, and approximating the function $g(n,L)$ as $1/2$ yields
\begin{align}
\left\langle t\right\rangle\approx 3\alpha_0 L_{x}L_{y}L_{z} = \frac{\alpha_0 V}{2D_0} ,\label{mfpt_asym}
\end{align}
where 
\begin{align}
\alpha_0=\int^{1}_{0}\int^{1}_{0}\int^{1}_{0}\frac{dxdydz}{\Omega_{0}(x,y,z)}\approx0.505 .\label{mfpt_app}
\end{align}
In order to see how good this approximation is we use the Euler-Maclaurin formula for the $p$'th order approximation of a sum
\begin{align}
&\sum^{L}_{n=1}f(n)=\int^{L}_{0}f(x)dx+\frac{f(L)-f(0)}{2} \nonumber\\
&+\sum^{p}_{k=1}\frac{B_{2k}}{\left(2k\right)!}\left[f^{(2k-1)}(L)-f^{(2k-1)}(0)\right]+R_{p} ,\label{euler}
\end{align}
where $B_{2k}$ are the Bernoulli numbers, $f^{(k)}(x)$ is the $k$'th derivative of $f(x)$, and $R_{p}$ is an error term. \textcolor{black}{Note that $p$ here is an arbitrary positive integer.} Using Eq.~(\ref{euler}) for each of the three sums in Eq.~(\ref{mean_fpt}), we find that the error term is of order $L^{-(2p+1)}_{J}$ for $J=x,y,z$. Next, note that the summand is such that the function $f(x)$ in Eq.~(\ref{euler}) satisfies $f(x)=f(L-x)$. Therefore, the second and third terms on the right hand side of Eq.~(\ref{euler}) are identically zero. Hence, we conclude that the difference between the exact MFPT, Eq.~(\ref{mean_fpt}), and the approximation, Eq.~(\ref{mfpt_app}), is smaller than $L^{-(2p+1)}_{x,y,z}$ for all $p$. This means that the error is exponentially small in $L_{x,y,z}$.

\section{1D searching}\label{1D}

In this appendix, we obtain the FPT \textcolor{black}{probability density} of a 1D random walk problem by solving the discrete diffusion equation. Due to the periodic boundary conditions of our model, the topology of the channel can be understood as a ring-like structure with a circumference of $\mathscr{L}_y$ and a single target that can be reached by the tracer molecule either from the left or from the right side. Let $P(y,t)$ denote the probability of finding the particle at a time $t$ at a position $y$, given that it has not been absorbed either by the target at the left edge, nor by the one at the right edge of the channel. Then, $P(y, t)$ obeys the discrete diffusion equation, Eq.~(\ref{eveq}), with the boundary conditions
\begin{equation}
P(0,t) = P(\mathscr{L}_y,t) = 0,
\end{equation}
Imposing the boundary and initial conditions on the general solution, Eq.~(\ref{ansatz3}), yields
\begin{align}
P(y,t)=\frac{2}{\mathscr{L}_{y}}\sum^{\mathscr{L}_{y}-1}_{n=1}e^{-\omega\left(k_{n}\right)t}\sin\left(k_{n}y_{0}\right)\sin\left(k_{n}y\right) .\label{P1Diff}
\end{align}

The first passage time to the target is related to the survival probability that the tracer did not yet reach the target at time $t$, $\mathcal{H}(t)$, by~\cite{Klafter}: 
\begin{equation}
F(t) = - \frac{\partial  \mathcal{H}(y,t)} {\partial t}.
\end{equation}
The survival probability is equal to the probability that the tracer diffusing on the channel of length $\mathscr{L}_{y}$ with absorbing boundary conditions, remains in the system:
\begin{equation}
\mathcal{H}(t) =  \sum_{y=1}^{\mathscr{L}_{y}-1} P(y,t).
\end{equation}
Thus, assuming that the particle starts its diffusive process form the middle of the channel, i.e., $y_0=\mathscr{L}_{y}/2$, we get that the FPT \textcolor{black}{probability density} is given by:
\begin{align}
& F(t) = \frac{2}{\mathscr{L}_{y}}\sum^{\mathscr{L}_{y}/2}_{m=1}\omega\left(k_{2m-1}\right) e^{-\omega\left(k_{2m-1}\right)t} \nonumber\\
& \times \left(-1\right)^{m+1}\cot\left(\frac{\pi(2m-1)}{2\mathscr{L}_{y}}\right) .
\end{align}
The MFPT is given by
\begin{align}
&\left\langle t\right\rangle=\int^{\infty}_{t}F(t)dt \nonumber\\
&=\frac{2}{\mathscr{L}_{y}}\sum^{\mathscr{L}_{y}/2}_{m=1}\left(-1\right)^{m+1}\frac{\cot\left(\frac{\pi(2m-1)}{2\mathscr{L}_{y}}\right)}{\omega\left(k_{2m-1}\right)} =\frac{3}{4}\mathscr{L}^{2}_{y} .
\end{align}
The same result can be obtained by setting $L_{x}=L_{z}=1$ in Eq.~(\ref{mean_fpt}).
The MPFPT is found by solving
\begin{align}
& \left.\frac{\partial F(t)}{\partial t}\right|_{t=t^{\ast}}=\frac{2}{\mathscr{L}_{y}}\sum^{\mathscr{L}_{y}/2}_{m=1}\omega^{2}\left(k_{2m-1}\right) e^{-\omega\left(k_{2m-1}\right)t^{\ast}} \nonumber\\
& \times \left(-1\right)^{m+1}\cot\left(\frac{\pi(2m-1)}{2\mathscr{L}_{y}}\right) =0,
\end{align}
which in the limit $\mathscr{L}_{y}\rightarrow\infty$ yields Eq.~(\ref{betadef}).

\section{Evaluation of the TSDT}
\label{TSDT}
The Laplace transform of the FPT \textcolor{black}{probability density}, $\tilde{F}(s)$, is finite for all values of $s$ with a non-negative real part, and all its poles have a negative real part. Since the number of poles is finite, the asymptotic behaviour of the FPT \textcolor{black}{probability density} is exponential $\exp\left(-t/\hat{t}\right)$. The TSDT, $\hat{t}$, is given by
\begin{align}
\hat{t}=1/\hat{s} ,
\end{align}
where $\hat{s}$ is the pole of $\tilde{F}(s)$ with the smallest real part (in absolute value). The poles of $\tilde{F}(s)$ are found by equating its denominator to zero, and thus $\hat{s}$ is the smallest root of $G(s)$, defined by
\begin{align}
&G(s)=\sum^{L_{x}-1}_{n_{x}=0}\sum^{L_{z}-1}_{n_{z}=0}\sum^{L_{y}-1}_{n_{y}=0}\frac{g(n_{x},x_{0},L_{x})g(n_{z},z_{0},L_{z})}{s+\omega\left(\frac{\pi n_{x}}{L_{x}}\right)+\omega\left(\frac{\pi n_{z}}{L_{z}}\right)+\omega\left(\frac{2\pi n_{y}}{L_{y}}\right)} .
\end{align}
Assuming that $\left|\hat{s}\right|\ll1$, i.e., $N \approx \sqrt{L_y}$ and large, $G(s)$ may be expanded around $s=0$ such that
\begin{align}
G\left(s\ll1\right)\approx\frac{1}{s}+\sum_{n_{x},n_{y},n_{z}}\frac{g(n_{x},x_{0},L_{x})g(n_{z},z_{0},L_{z})}{\omega\left(\frac{\pi n_{x}}{L_{x}}\right)+\omega\left(\frac{\pi n_{z}}{L_{z}}\right)+\omega\left(\frac{2\pi n_{y}}{L_{y}}\right)} ,
\end{align}
and thus $\hat{s}$ may be approximated by
\begin{align}
\hat{s}\approx \left[\sum_{n_{x},n_{y},n_{z}}\frac{g(n_{x},x_{0},L_{x})g(n_{z},z_{0},L_{z})}{\omega\left(\frac{\pi n_{x}}{L_{x}}\right)+\omega\left(\frac{\pi n_{z}}{L_{z}}\right)+\omega\left(\frac{2\pi n_{y}}{L_{y}}\right)}\right]^{-1} .
\end{align}
For large enough $L_{y}$, the sum can be approximated by the MFPT, and thus $\hat{t} \approx\left\langle t\right\rangle$. \textcolor{black}{For a similar derivation of the TSDT in more general systems see Ref.~\cite{Hartich2018}.}

\end{document}